\documentclass[useAMS,usenatbib]{mn2e}

\usepackage{graphicx}
\usepackage{times}
\usepackage{natbib}

\newcommand{\apjs}{ApJS}

\newcommand{\apjl}{ApJL}

\newcommand{\aap}{A \& A}

\newcommand{\enzo}{\it{\small ENZO}}

\begin{document}

\title[CRs in large-scale structures] {Simulations of cosmic rays in large-scale structures: numerical and physical effects}
\author[F. Vazza, C. Gheller, M. Br\"{u}ggen]{F. Vazza$^{1,2}$\thanks{E-mail: franco.vazza@hs.uni-hamburg.de}, C. Gheller$^{3}$, M. Br\"{u}ggen$^{1}$\\
$^{1}$ Hamburger Sternwarte, Gojenbergsweg 112, 21029 Hamburg, Germany\\
$^{2}$INAF/Istituto di Radioastronomia, via Gobetti 101, I-40129 Bologna,
Italy\\
$^{3}$  CSCS, Via Trevano 131, CH-6900 Lugano, Switzerland}

\date{Received / Accepted}
\maketitle
\begin{abstract}

Non-thermal (relativistic) particles are injected into the cosmos by structure formation
shock waves, active galactic nuclei and stellar explosions. 
We present a suite of unigrid cosmological simulations (up to $2048^3$) using a two-fluid model in the grid code {\enzo}. The simulations include the dynamical effects of cosmic-ray (CR) protons and cover a range of theoretically motivated acceleration efficiencies. 
For the bulk of the cosmic volume the modelling of CR processes is rather stable with respect to resolution, provided that a minimum (cell) resolution of $\approx 100  ~\rm kpc/h$ is employed. However, the results for the innermost cluster regions depend on the assumptions for the baryonic physics. Inside clusters, non-radiative runs at high resolution tend to produce an energy density of  CRs that are below available upper limits from the FERMI satellite, while the radiative runs are found to 
produce a higher budget of CRs. We show that weak ($M \leq 3-5$) shocks and shock-reacceleration are crucial to set the level of CRs in the innermost region of clusters, while in the outer regions the level of CR energy is mainly set via direct injection by stronger shocks, and is less sensitive to cooling and feedback from active galactic nuclei and supernovae.

\end{abstract}

\label{firstpage} 


\section{Introduction}
\label{sec:intro}

The intergalactic medium consists of a mixture of thermal particles and relativistic particles (or cosmic ray, hereafter CR), coupled to ubiquitous magnetic fields and turbulent flows. The hierarchical process of structure formation proceeds via virialization of kinetic energy during the gravitational infall of gas. On $\sim \rm Mpc$ scales, this process occurs via mergers and supersonic infall of cold matter.
The resulting shock waves and turbulence are crucial for the 
amplification of magnetic fields and for the acceleration of CR from the thermal pool.\\
Evidence for non-thermal particles in galaxy clusters is manifold. Observations in the radio band \citep[e.g.][for recent reviews]{fe08,fe12} showed the presence of $\sim 0.1-10 \mu G$ magnetic fields and $\sim \rm GeV$ CR electrons in tens of galaxy clusters.
Unlike in the Milky Way, where an approximate equipartition between thermal and non-thermal energy has been suggested \citep[e.g.][]{2005AN....326..414B}, in the case of the interior of galaxy clusters there is evidence that non-thermal energy is limited to a few percent of the thermal energy. This estimate comes from the absence of hadronic $\gamma$-ray emission from the intracluster medium (ICM), as reported from a number of recent high-energy observations, such as the ones of Abell 85 by {\it HESS} \citep[][]{dom09}, of the Perseus cluster by {\it MAGIC} \citep[][]{alek10,alek12}, of the Coma cluster by {\it VERITAS} \citep[][]{arl12} and of a local sample of clusters by {\it FERMI} \citep[][]{ack10,2013arXiv1308.6278H,fermi13}.\\
Once accelerated, CR protons are rather unaffected by energy losses and spatial diffusion, which have a typical life-time comparable to or larger than the Hubble time  \citep[][]{bbp97}.  Thus they can collide with the thermal protons of the ICM, eventually producing $\gamma$-radiation from the decay of neutral pions \citep[e.g.][]{2003MNRAS.342.1009M,bl07,2011PhRvD..84l3509P}.\\
The constraints on the energy density in CR protons from the non-detection of $\gamma$-ray emission are on the level of a few percent of the thermal gas energy for most of the observed (nearby) clusters, and rather independent of their dynamical state \citep[][]{ack10,alek12}. \\ 
In addition to that, secondary particles are continuously injected
into the ICM via proton--proton collisions, possibly leading to
detectable synchrotron radiation \citep[e.g][]{bl99, de00}. 
The detection or non-detection of non-thermal emission from cluster
centres may inform us about the energy budget of non-thermal
particles within galaxy clusters. 
The combined analysis of radio observations and $\gamma$-ray upper limits suggests that most of the observed large-scale
radio emission in clusters cannot be due to secondary electrons released in hadronic
collisions \citep[][]{br07,donn10,2009A&A...508..599B,jp11,2012MNRAS.426..956B}.\\
The presence of a non-thermal pressure component (of the order of $\leq 10$ percent of the thermal pressure support over the full cluster volume) has been suggested by a number of studies comparing the hydrostatic and the gravitational mass in X-ray observed clusters \citep[e.g.][]{2008MNRAS.388.1062C, 2013A&A...551A..22E,2013SSRv..tmp...51E}.  Whether this non-thermal support comes in the form of magnetic fields, CR protons or turbulent motions is still under investigation. 
\bigskip

A very relevant mechanism to accelerate CRs in large-scale structure is diffusive shock acceleration (DSA) in structure formation shocks \citep[e.g.][] {be78, bo78, dv81, ebj95, kj90, md01,kj07,ca10}.
In the  "test-particle" regime, where the accelerated particles do not have any dynamical effects on the 
structure of the shock, DSA predicts the energy spectrum of $N(p) dp \propto p^{-q} dp$ for the
accelerated particles, where $q$ is related to the Mach
number via $q=2 (M^{2}+1)/(M^{2}-1)$. 
However, models of DSA predict that
at strong shocks, $M>5$, the fraction of accelerated CR
particles becomes non-negligible compared to the thermal pool and that a significant fraction of the shock
energy is transferred to the CR particles, which in turn modifies
the shock structure.  At scales of the order of a few hundreds of the proton gyro-radii, the diffusion of accelerated 
particles across the shock is expected to produce a shock-precursor, which modifies the
overall compression produced by the entire shock structure in a non-linear way \cite[e.g.][]{dv81,bl04,ab05}. The result is a larger shock compression and 
an amplification of the magnetic fields in the pre-shock
region \citep[e.g][for recent reviews]{2012JCAP...07..038C,2012JKAS...45..127K}.
While the basic
predictions of DSA have been tested successfully with multi-band observations of radio and $\gamma$-ray emission from 
remnants of supernovae \citep[SNe, e.g.][]{rey08, vi10, ed11}, a more detailed 
analysis of SN remnants suggests that some modification of the
theory is necessary because the observed high-energy spectra of accelerated particles show evidence
of a {\it convex} spectrum, contrary to the results of \citet[][]{2012arXiv1206.1360C}. \\
Also for the regime of Mach numbers more pertinent to the ICM, $M<10$ \citep[e.g.][]{quillis98,ry03,pf06,sk08,va09shocks,va11comparison,2013MNRAS.428.1643P}, the efficiency of DSA is not yet robustly constrained by theory due to the difficulty of modelling the large range of
spatial and temporal scales involved in the diffusive acceleration at such shocks. More recently, 
several groups employing particle-in-cells methods investigated
additional mechanisms for proton and electron acceleration at shocks, such as shock drift acceleration,
\citep{2012ApJ...744...67G,2011ApJ...733...63R,2012ApJ...755..109M}. However, it is not trivial to relate these studies
to ICM shocks.\\

In addition to structure formation shocks,  active galactic nuclei (AGN)  can inject non-thermal
particles into the ICM through direct injection at the jet base 
\citep[e.g.][]{mb07,2008MNRAS.387.1403S,guo08,ma11,fuj12} or as a result of the 
shock waves excited by the interplay between the AGN outflow
and the surrounding ICM \citep[e.g.][and references therein]{va13feedback}.\\
Finally, the cumulative effect of multiple supernovae in cluster galaxies is expected to trigger
galactic winds that release CRs into the ICM.
\citep[][]{1996SSRv...75..279V,Volk&Atoyan..ApJ.2000}.\\
The inclusion of CR-processes in cosmological simulations was first presented by \citet[][]{mi01}, who performed fixed-grid simulations of large-scale structure formation that include the acceleration, transport and energy losses of the high energy particles. CRs were injected at shocks according to the thermal leakage model and then accelerated to a power-law distribution as predicted by the test particle limit of the DSA theory. The dynamical back reaction of accelerated CRs on the thermal gas was, however, not included in this early work.\\\citet{pf07} presented smoothed-particle-hydrodynamics (SPH) simulations that allowed for the pressure feedback of accelerated CR particles on the thermal gas, reaching a high spatial resolution in the innermost regions of galaxy clusters.\\ 
These studies have suggested that a significant budget of non-thermal energy within clusters can also affect the expected scaling
behaviour of temperature, entropy and pressure with the host cluster mass,
limiting their use for precision cosmology.\\

In the recent past, we have developed a two-fluid model for CR-dynamics in the {\enzo} code \citep[][]{enzo13}. We have applied this code to study the pressure feedback from CRs in galaxy clusters making use of adaptive mesh refinement (AMR).
Our method enables us to model shock waves at high resolution even in the outer regions of clusters, and to include the dynamical feedback of CRs on the thermal gas \citep[][]{scienzo}. \\
In a follow-up paper, we implemented feedback from AGN within galaxy clusters, and monitored the hadronic $\gamma$-ray signature associated with each feedback
mode, placing some constraints on the power released from AGN inside clusters \citep[][]{va13feedback}.\\

In this paper we extend our previous work by performing large-scale simulations on uniform grids. In particular, we test several models for the
evolution of the gas and for the acceleration efficiency of CRs at shocks. 
In this first paper we focus on the large-scale properties of the gas and CR distributions across $\sim 7$ orders of magnitude in gas density and for different cosmic epochs. 
We monitor the distribution of shock waves and the CR energy budget resulting from different physical assumptions, and we assess the degree of convergence in these statistics.\\ 

The outline of this paper is as follows:
our numerical methods are described in Sec.\ref{sec:methods}, and then our results are presented in Sec.\ref{sec:res}. Sec.\ref{sec:conclusion} gives a critical discussion on the open physical and numerical issues on our modelling, and lists our main results. 

\bigskip

\section{Numerical methods}
\label{sec:methods}

{\enzo} is an AMR code using the Piecewise Parabolic Method (PPM) to solve the equations of hydrodynamics, 
originally written by \citet{br95}, and currently developed as a collaborative effort of scientists at many universities and national laboratories \citep[][]{enzo13}.  
{\enzo}  uses a particle-mesh N-body method (PM) to follow
the dynamics of the collision-less Dark Matter (DM) component 
\citep{he88}, and an adaptive mesh method for ideal 
fluid-dynamics \citep{bc89}.
The DM component is coupled to
the baryonic matter (gas) via gravitational forces, calculated from the 
total mass distribution (DM+gas) solving the Poisson equation with 
an approach based on the Fast Fourier Transform. In its basic version, the gas component is described as a perfect fluid ($\gamma=5/3$) and its dynamics
is calculated by solving conservation equations of mass, energy and momentum over a computational mesh,
using a Eulerian solver based on the 
Piecewise Parabolic 
Method (PPM, \citealt{cw84}). This scheme is 
a higher-order extension of Godunov's shock capturing
method \citep{go76}, and it is at least second--order accurate in space (up
to the fourth--order in 1--D, in the case of smooth flows and small time-steps) and
second--order accurate in time.\\

On the basis of the public 1.5 version of {\enzo}, we have implemented our methods for the evolution and feedback of CR particles  \citep[][]{scienzo}, as well as our implementation of energy release from AGN \citep[][]{va13feedback} and SNe. Unlike in our previous work \citep[e.g.][]{scienzo,va13feedback} here we did not make use of AMR in order to quantify the spatial distribution of CRs across all cosmic environments without being affected by variable resolution effects, particularly at shocks.

For the simulations presented here, we assumed a ''WMAP 7-year'' cosmology with
$\Omega_0 = 1.0$, $\Omega_{B} = 0.0455$, $\Omega_{DM} =
0.2265$, $\Omega_{\Lambda} = 0.728$, Hubble parameter $h = 0.702$, a normalisation for the primordial density power
spectrum $\sigma_{8} = 0.81$ and a spectral index of $n_s=0.961$ for the primordial spectrum of initial matter
fluctuations.  All simulations are started at $z_{\rm in}=30$.\\

\subsection{Radiative cooling and reionization}

Radiative cooling is modelled assuming a primordial composition of a fully ionized H-He plasma 
with a uniform metallicity of $Z=0.3 ~Z_{\odot}$ (where $Z_{\odot}$ is the solar metallicity). 
As input we use the tabulated APEC emission model \citep[e.g.][]{2001ApJ...556L..91S} in order to compute the cooling function 
of each cell at run-time and as a function of temperature.
For the cold gas in the simulated volume, with temperature $T \leq 10^4 \rm K$, we use the cooling curve of \citet{2011ApJ...731....6S}, which is derived from a complete set of metals (up to atomic number 30), obtained with the chemical network of the photo-ionization software {\it Cloudy} \citep[][]{1998PASP..110..761F}.\\
In order to model the UV re-ionization background  \citep[][]{hm96}, we enforced a temperature floor for the gas in the redshift range $4 \leq z \leq 7$, as discussed in  \citet{va10kp}. 
In a recent paper, \citet{2013ApJ...763...38S} showed that the use of metallicity-dependent gas cooling is essential to prevent early overcooling in {\small ENZO} runs, and that adjusting the amount of energy and metal feedback can have a significant impact on observable X-ray quantities of the gas. Considering that the best spatial resolution of our ensemble of large unigrid runs is
still moderate ($\geq 48$ comoving $\rm kpc/h$), we expect that any effect of overcooling in dense sub-halos at high redshift is alleviated by the lack of high spatial resolution. Furthermore, the heuristic model for energy feedback from AGN that we use in the runs with cooling is effective enough to prevent catastrophic cooling within halos.

\subsection{Cosmic ray physics}

\label{subsec:cr}

The methods to model the injection, advection and pressure feedback of CRs have already been introduced and tested in \citet{scienzo} and \citet{va13feedback}. We assume that CRs are injected at shocks with an acceleration efficiency, $\eta(M)$, that only depends on the Mach number, $M$, which is given by various prescriptions for DSA 
\citep[e.g.][]{be78, bo78, dv81, ebj95, kj90, md01,kj07,2012arXiv1206.1360C}. 
We follow the approach suggested by \citet{mi07} and treat the physical processes across the shock transition in a sub-grid fashion. The  details of the particle acceleration
mechanism are included in the solution of the Riemann problem and by imposing the additional pressure of CRs.
At each detected shock, new CR energy is {\it injected} in the system by integrating the energy flux through each shocked cell by the  time step and the cell surface:
the energy density of CRs is given by: 
\begin{equation}
E_{\rm cr}= \eta (M) \cdot \frac{\rho_{\rm u} v_{s}^{3}}{2} \cdot \frac{\Delta t_{\rm l}}{\Delta x_{\rm l}},
\label{eq:secondary}
\end{equation}
where $\rho_{\rm u}$ is the pre-shock density, $v_{\rm s}$  is the shock velocity, $\Delta t_{\rm l}$ and $\Delta x_{\rm l}$ are the time step and
the spatial resolution, respectively.  To 
ensure energy conservation, the thermal
energy in the post-shock region is {\it reduced} proportionally at run-time {\footnote{We do not model the escape of high-energy CRs out of the shock region, whose size is set by the cell size. Otherwise, we should model a non-local scattering of the escaping particles far from the shock region, a condition which cannot easily be treated with a two-fluid model. No simple physical picture of this mechanism is available because the fraction of escaping particles depends on the CR-driven amplification of the magnetic field in the shock region in a non-linear way.  We refer the reader to recent works of \citet{2013MNRAS.431..415B} and \citet{2013MNRAS.tmp.2295B} for reviews on this topic. Thus we assume that all the dissipated shock energy is redistributed locally in the system, either in the form of thermal or CR energy.}}.\\

\begin{figure}
\includegraphics[width=0.45\textwidth]{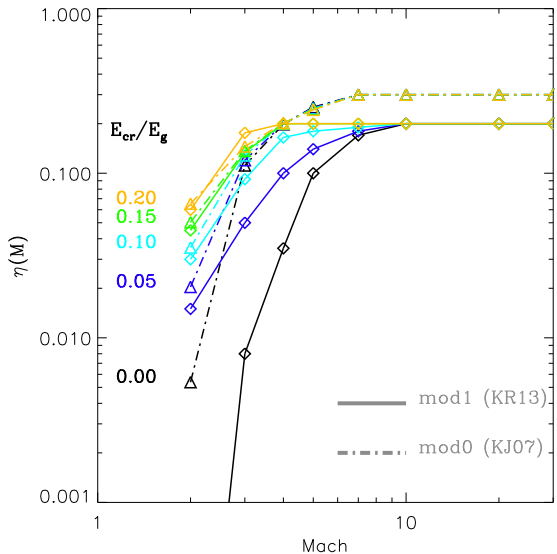}
\caption{Acceleration efficiency of CRs as a function of Mach number adopted in our runs. The dot-dashed lines show the acceleration model of Kang \& Ryu (2013), while the solid lines show the acceleration model of \citet{kj07}. The different colors show the increased acceleration efficiency in case of an increasing energy ratio of pre-existing CRs in the pre-shock cell.}
\label{fig:eta}
\end{figure}

In this work, we used two acceleration efficiencies for CR protons at shocks: one by \citet{kj07}, that we already used in our previous works \citep[][]{scienzo,va13feedback}, and a newer version with a reduced acceleration efficiency at $M>10$, presented in \citet{kr13}. In this second model, the effect of the magnetic field amplification by CR-driven instabilities has been included in a parametric way, yielding an increase of the Alfv\'{e}nic drift of scattering centres across the shock, and to a generally lower acceleration efficiency at strong shocks. Figure \ref{fig:eta} shows the acceleration efficiency as a function of $M$ in the two models, with and without a pre-existing population of seed CRs. In both cases the maximum acceleration efficiencies at strong
shocks were $\eta(M)=0.3$ in \citet{kj07} and $\eta(M)=0.2$ for \citet{kr13}.\\
The injection of CRs at shocks is switched on whenever the shocks run over a large enough overdensity,
$\geq \rho_{\rm cr}$ (where $\rho_{\rm cr}$ is the cosmological critical gas density). The choice of this minimum density is somewhat arbitrary, and ultimately depends on the assumption of a minimum suitable
ambient magnetic field for DSA to work. Our assumption here is that at least a pre-shock value of magnetic field in the range $B \sim 10^{-4}-10^{-3} \mu G$ is necessary for DSA to work, and follows from the average scaling of magnetic field with gas density, as suggested by cosmological MHD 
simulations \citep[e.g.][]{do08,sk13}. Our tests in \citet{scienzo} have shown that this choice 
is not crucial for the properties of CRs within large-scale structures at low redshift, but it
can change the budget of CRs at higher redshifts. If a much larger value of the critical
overdensity is chosen, the injection of CRs starts at later times.\\

After the injection of CRs, the total effective pressure is
$P_{\rm eff}=P_{\rm g}+P_{\rm cr}=\rho [(\gamma-1)e_{\rm g}+(\gamma_{\rm cr}-1)e_{\rm cr}]$, where $\gamma=5/3$, $\gamma_{\rm cr}=4/3$, $e_{\rm g}$ is the gas energy per mass and $e_{\rm cr}$ is the CR energy per mass. 
The dynamical feedback of CR pressure is treated in the Riemann solver by
updating the gas density fluxes in the 1--D sweeps along the coordinate axes and using the (pressure-weighted) effective
gamma factor ($\gamma_{\rm eff}=\frac{(\gamma P_{\rm g}+\gamma_{\rm cr} P_{\rm cr})}{P_{\rm g}+P_{\rm cr}}$), for the computation of the local sound speed in cells \citep[][]{mi07}.

As in \citet{scienzo}, we fix the relativistic value of $\gamma_{\rm cr}=4/3$ everywhere, which corresponds to the flattest possible momentum spectrum
of CRs, through $\gamma_{\rm cr}=q/3$ (with $f(p) \propto p^{-q}$ and $q=4$ for $\gamma_{\rm cr}=4/3$). This is reasonable since once the CR energy density is specified, the CR pressure depends only weakly on the spectral shape of $f(p)$ and on the cut-off momentum \citep[e.g.][]{1993ApJ...402..560J,ju08}.\\

To model the injection of CRs at each shocked cell, we rely on a run-time shock finder 
method based on 3--D pressure jumps, that we already introduced in \citet{scienzo}. At each time step, we flag cells
with a negative divergence, $\nabla \cdot {\vec v} < 0$, whose local gradients of temperature and entropy satisfy $\nabla S \cdot \nabla T>0$ \citep[][]{ry03}. Then the pre- and the post-shock regions are assigned based on
the gradient of the gas temperature
and entropy. Finally, the local shock Mach number is computed by inverting the shock jump condition for gas pressure. In \citet{scienzo} we have already shown that this
method qualitatively yields results similar to a more elaborate (and time consuming) shock finding method based on the analysis of 3-D velocity jumps \citep[][]{va09shocks}.

\subsubsection{Shock re-acceleration}

For weak shocks ($M \leq 3-4$) the effect of shock re-acceleration is as important as the direct injection in DSA, and more significant than adiabatic compression of CRs
\citep[][]{kr13,2013MNRAS.435.1061P}. This can be particularly relevant in the case of shocks caused by AGN feedback and in internal merger shocks, where at late redshifts the ICM is already enriched with
CRs.\\
The presence of CRs in the pre-shock region is treated using a different analytical function for $\eta(M)$ \citep[][]{kj07,kr13},
dependent on the ratio $E_{\rm cr}/E_{\rm g}$ in the pre-shock (where $E_{\rm cr}=\rho e_{\rm cr}$ and $E_{\rm g}=\rho e_{\rm g}$). In our case, we calculate $\eta(M)$ using a linear interpolation between the bracketing cases of $E_{\rm cr}/E_{\rm g}=0$ and $E_{\rm cr}/E_{\rm g}=0.2$. 
The post-shock thermal gas energy is reduced at run-time accordingly, as in \citet{scienzo}.

\subsubsection{Hadronic and Coulomb losses}
\label{subsubsec:hadronic}

Cosmic rays can lose energy via
binary interactions with thermal particles of the ICM. This channel
of energy exchange between thermal and relativistic particles
in the ICM is important for the high gas 
density ($\rho/(\mu m_{\rm p}) \geq 10^{-2} \rm cm^{-3}$) of cool
cores. Relativistic protons transfer energy to the thermal gas via Coulomb collisions with the ionized gas.
They can also interact hadronically with the ambient ICM and produce mainly pions, provided their kinetic energy exceeds the threshold of 282
MeV for the reaction. While the charged pions decay into $e^{\pm}$ and neutrinos, the neutral pions decay after a mean lifetime of $\approx 9 \cdot 10^{-17}$ s into $\gamma$-rays.
To estimate the total energy transfer rate between CRs and thermal
gas in both mechanisms, we need to determine
the CR energy spectrum. Since this information is not readily available in the two-fluid model, we must assume an
approximate steady-state spectrum for the CR energy distribution.
We set a spectral index of $\alpha=2.5$ for the particle energy, which is a reasonable
average value over the overall population of cosmic shocks \citep[][]{2010MNRAS.409..449P,scienzo}, and we computed the total Coulomb and hadronic loss rates as a function of the ICM density and of $E_{\rm cr}$ for
each cell, as in \citet{guo08}:

\begin{equation}
\Gamma_{\rm coll}=-\zeta_{\rm c} \frac{n_{\rm e}}{\rm cm^{-3}} \frac{E_{\rm cr}}{\rm erg \cdot cm ^{-3}} \rm erg \cdot s^{-1} \cdot cm^{-3},
\label{eq:gamma_coll}
\end{equation}

where $n_{\rm e}$ is the electron number density,
and $\zeta_{\rm c}= 7.51 \cdot 10^{-16}$ is the coefficient for all collisional energy loss terms. In hadronic
collisions, only $\sim 1/6$ of the inelastic energy goes into secondary electrons
\citep[][]{2004A&A...413..441C,guo08}. The bulk of CR electrons
($\gamma \sim 10^{2}$) will heat the ICM through Coulomb interactions, plasma oscillations and excitation of Alfv\'{e}n waves
\citep[e.g.][]{guo08}. Therefore,  we can 
assume that these secondary electrons lose most of their energy
through thermalization and thus heat the ICM. Similar to Eq.\ref{eq:gamma_coll}, the heating rate of the ICM through Coulomb
and hadronic collisions can be computed as:

\begin{equation}
\Gamma_{\rm heat}=\xi_{\rm c} \frac{n_{\rm e}}{\rm cm^{-3}} \frac{E_{\rm cr}}{\rm ergs \cdot cm^{-3}} \rm erg \cdot s^{-1} \cdot cm^{-3}, 
\label{eq:gamma_heat}
\end{equation}

where $\xi_{\rm c}=2.63 \cdot 10^{-16}$ \citep[][]{guo08}.
In our simulations with radiative cooling and AGN feedback, the rate
of energy loss due to these collisions is small, typically $\sim 10^{-3}-10^{-4}$ of $E_{\rm cr}$ or $e_{\rm g}$ during the time step. This allows us to use a simple first-order integration to compute the energy losses of CRs (and the corresponding gas heating rate) at run-time. \\

In a recent paper, \citet{2013MNRAS.432.1434F} studied the impact of assuming slightly different
choices of average spectra (from $\alpha=2$ to $\alpha=3$ in the proton spectra) and minimum energy for the CR protons (from $p_{\rm p,min}c=43$ MeV to $p_{\rm p,min}c=440$ MeV ) in the evolution of
single-object simulations of clusters. They found negligible differences in the
final evolution of the simulated ICM. 

Our previous work showed that modelling these processes at run-time decreases
the CR energy by a factor $\sim 10$ only within cluster cores, while they yield identical results for the remaining cluster volume compared to runs that neglect losses \citep[][]{va13feedback}.

\subsection{Sources of energy feedback}

\subsubsection{Active galactic nuclei - bipolar thermal feedback}
\label{subsubsec:agn}

In \citet{va13feedback} we introduced our simple treatment of (thermal) energy release from AGN within galaxy clusters, that we use in this study for 
a subset of our runs.
This is necessary because, in absence of a powerful source
of non-gravitational energy in the dense ICM, most structures
would undergo catastrophic cooling.
In our simplified approach, we trigger an ``AGN-feedback'' event wherever $\rho \geq \rho_{\rm min}$
in a cell, where $\rho_{\rm min} \approx 10^{-2}  \rm cm^{-3}$  \citep[][]{2007MNRAS.380..877S,teyssier11,2012MNRAS.422.3081M}. We also require that if multiple cells above $\rho_{\rm min}$
are found in close contact, only the densest within a $(1 ~\rm Mpc/h)^3$ region is used. 
This method by-passes, both, the
problem of monitoring the mass accretion rate onto the central black hole (BH)  within
the galaxy and the complex and small-scale physical processes which are supposed
to couple the energy from the BH to the surrounding gas (i.e. the launching of strong winds due to the radiation pressure of photons from the accretion disc). This is unavoidable, given that our best resolution is orders of magnitude larger than the accretion disc region, let alone the difficulty of modelling the radiative transfer of photons from the accretion region.\\
In the following, we will refer to the cells exceeding this density threshold
and powering energy feedback as ''AGN-cells''.
Based on our tests in \citet{va13feedback} we start feedback at $z_{\rm AGN}=4$, and we impose an energy release of $E_{\rm AGN} \sim 10^{59} \rm erg$ during the timestep, which typically corresponds to an initial temperature
around our AGN-cell of $\sim 5 \cdot 10^{7}-5 \cdot 10^{8} \rm K$ at the injection burst. 
Contrary to our previous work, here we employ only one mode of AGN feedback. We make use of a ``bipolar thermal feedback'', meaning 
that thermal energy is released into the ICM along two cells at the opposite side of the AGN-cell.
The direction of the two jets is randomly selected along one of the
three coordinate axes of the simulation.\\
We prefer the use of this scheme, instead of a kinetic one \citep[e.g.][]{dubois10,gaspari12}, or of other schemes \citep[e.g.][]{short12,2012MNRAS.427.1614Y} because the conversion of kinetic into thermal energy is expected to take place at a scale of a few tens of $\rm kpc$ from the AGN, typically of the order of (or below) our resolution \citep[][]{2009MNRAS.395.2317P}. 

\subsubsection{A population of high-redshift supernovae}
\label{subsubsec:sn}

Our runs neglect star formation and a detailed treatment of energy feedback
from SNe and stellar winds.
Hydrodynamical simulations suggest that while 
SNe are important to reproduce the observed metal
distribution of the ICM, they do not have a significant impact on the
thermal history of the ICM on large scales \citep[e.g.][]{short12}. 
However, in one re-simulation we tested the maximal impact of the injection
of CR by SNe with this basic treatment, and simulated the integrated effect of a population of SNe inside each halo, which we select with the thresholds of $\rho/\rho_{\rm cr} \geq 50$ and $T>10^7 \rm K$ in the simulation. Then we increase the gas thermal energy
by $E_{\rm SN}=\alpha_{\rm SN} \cdot E_{\rm g}$ in a very sharp transition peaked at 
redshift $z_{\rm SN}$.  The CR energy budget within the same cells is then increased by 
$0.3 \cdot E_{\rm SN}$, which we set following \citet{1996SSRv...75..279V}. This way, we can obtain a rough estimate of the integrated dynamical effect played by the thermal and CR output of SNe at low redshift ($z<1$). Following the work of \citet{1996SSRv...75..279V} we explored values in the range $\alpha_{\rm SN}=0.1-0.4$ and $z_{\rm SN}=2-4$, resulting in very small differences in the final CR and thermal energy budgets at low redshift. In this paper, we will show only results for the choices of $\alpha_{\rm SN}=0.25$ and $z_{\rm SN}=2$.\\
Contrary to other studies \citep{2012A&A...540A..77D,2013arXiv1307.6215S}, we assume an initial budget of CR energy which does not dominate over the thermal gas, at least on
the scale of the host galaxies.\\
In most cases, at injection this feedback mechanism does not trigger shocks within clusters, but instead leads to adiabatic expansion and, later, to an outflow from the host halo.
Our implementation produces a typical outflow velocity a few $\sim 10^2 \rm Myr$ after the injection of the order of $\sim 100-300 ~ \rm km/s$ outside the virial
volume of proto-halos, which in most cases is larger than the escape velocity of each halo at that redshift.
This value for the outflow velocity is in qualitative agreement with real observations of high-redshift compact galaxies \citep[e.g.][]{2002Ap&SS.281..461P,2002ApJ...570...92D, 2006ASPC..353..363P} and in line with other
numerical simulations \citep[e.g.][and references therein for an updated review]{2013MNRAS.430.3213B} on the same scale.

\begin{figure*}
\includegraphics[width=0.97\textwidth]{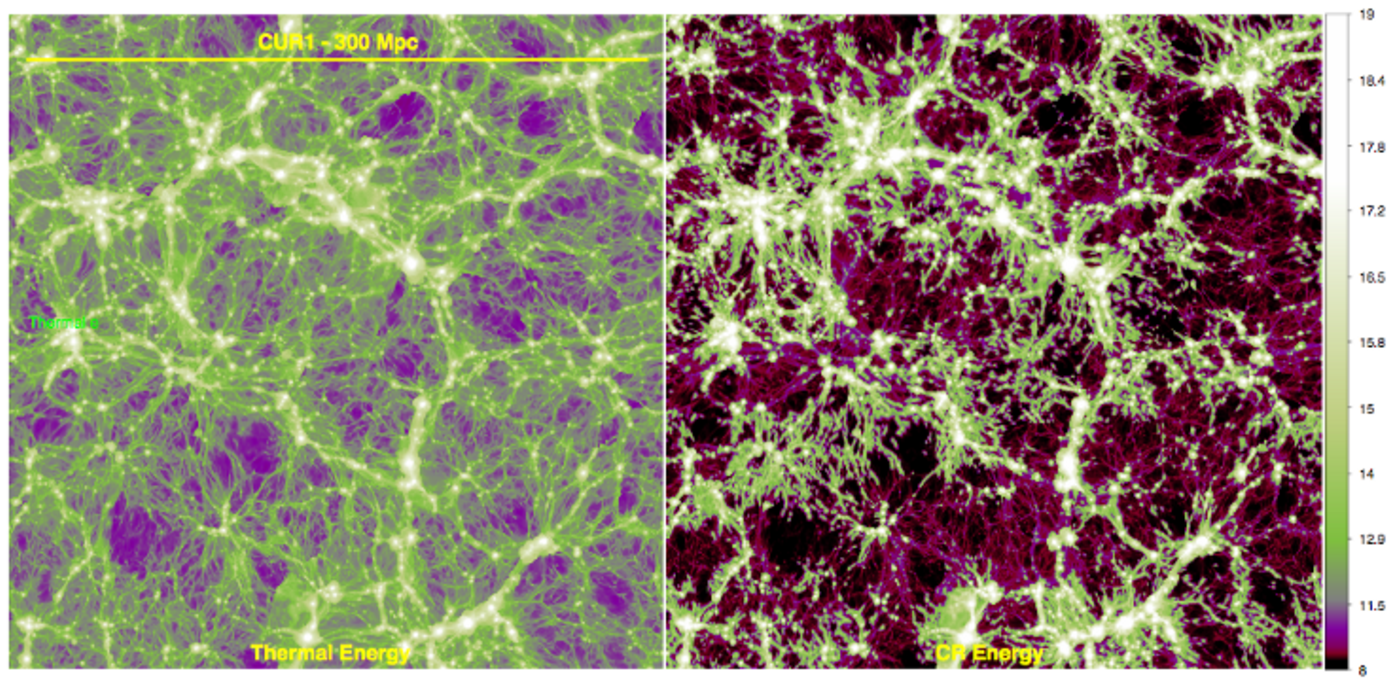}
\includegraphics[width=0.97\textwidth]{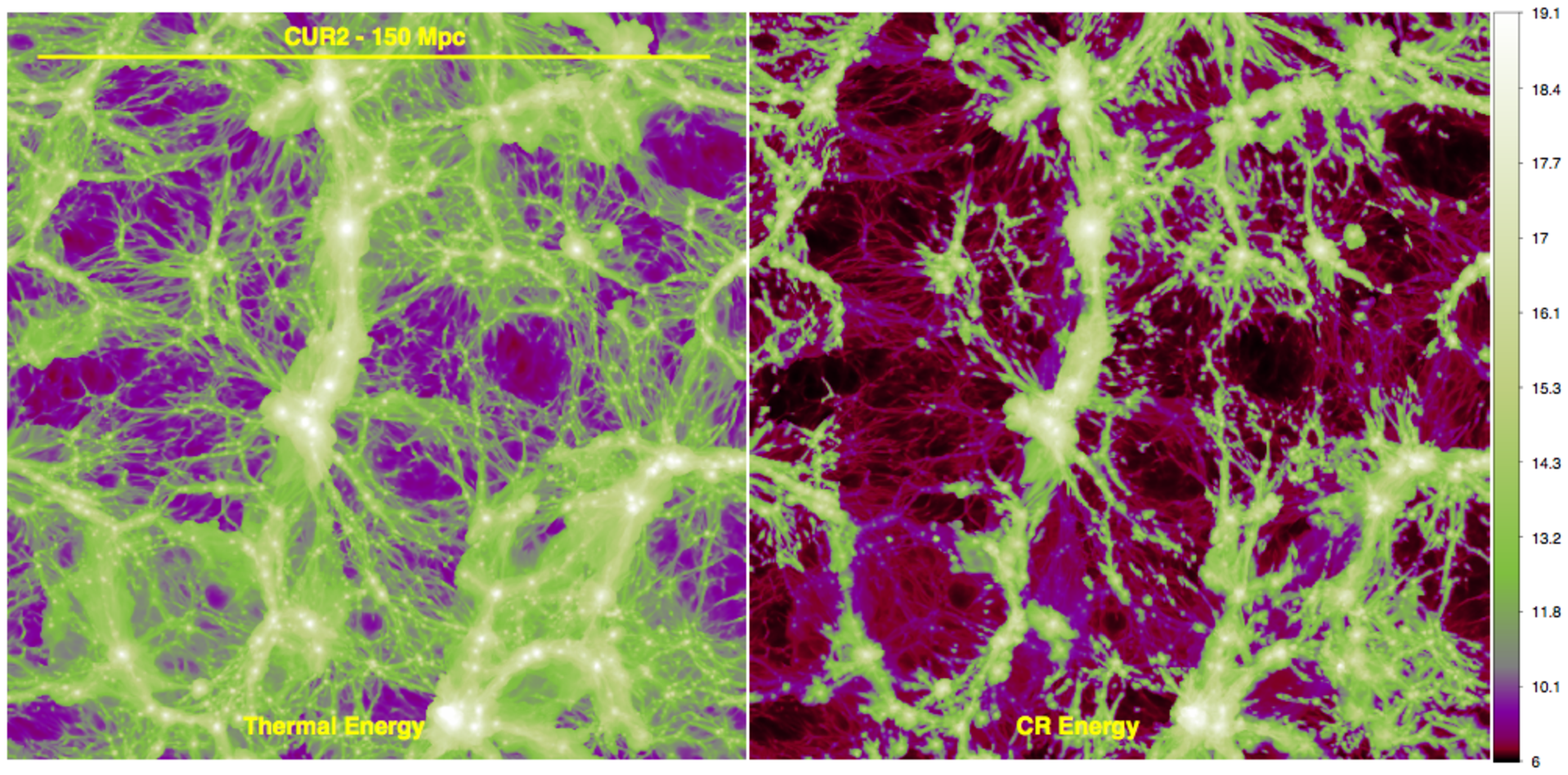}
\caption{Maps of gas energy (left panels, in $\log_{\rm 10}E_{\rm g}$, arbitrary units) and of CR energy (right panels, in $\log_{\rm 10}E_{\rm g}$, arbitrary units)  for the two non-radiative $1024^3$ CUR1\_1024\_1 and CUR2\_1024\_1, employing the low efficiency CR acceleration model \citep[][]{kr13}. Each image is $L \times L$, where $L$ is 216 and 108 $\rm Mpc/h$, respectively, while each map is the projection along a line of sight of  $\approx L/8$.}
\label{fig:maps}
\end{figure*}

\begin{figure*}
\includegraphics[width=0.97\textwidth]{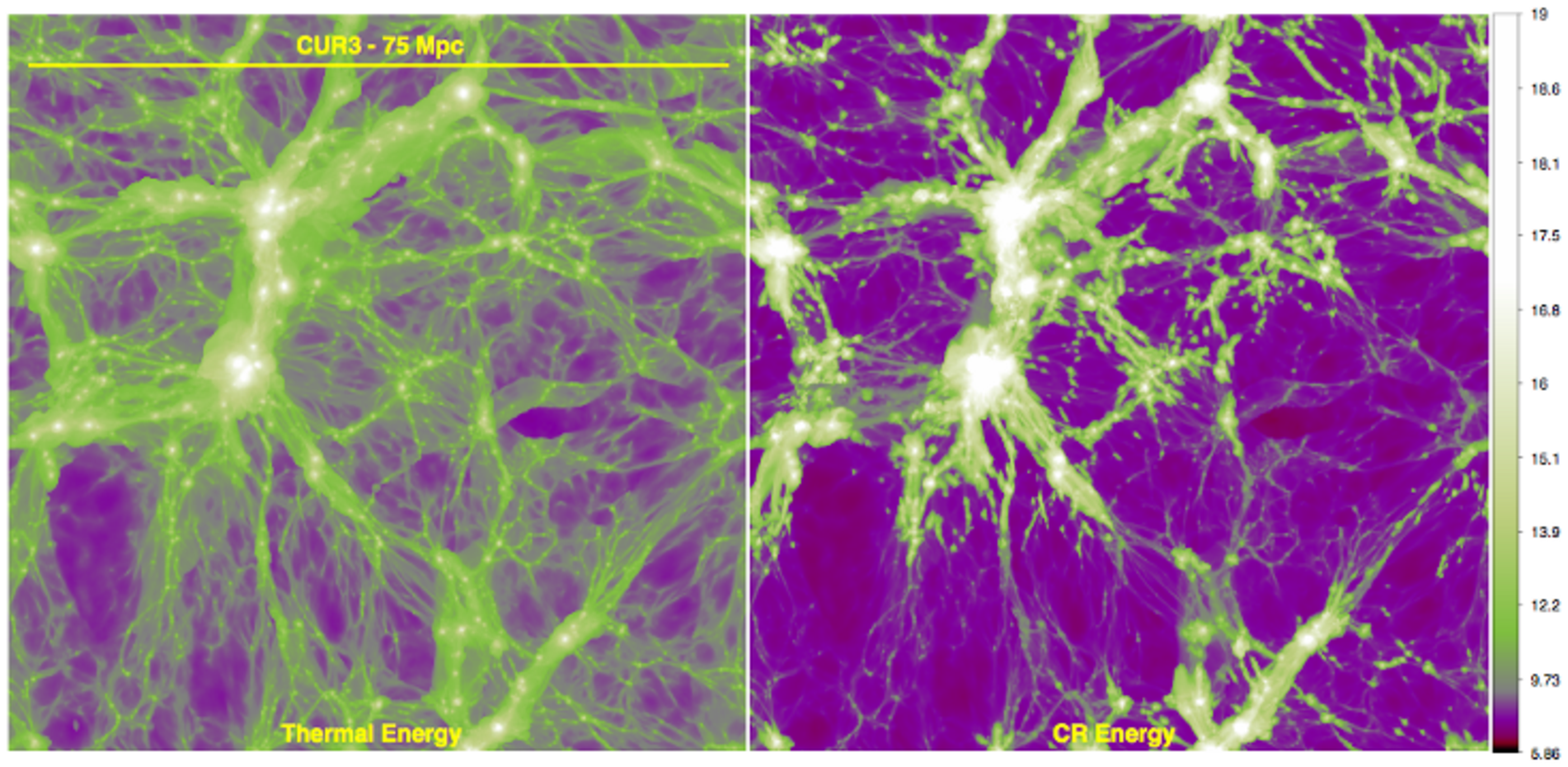}
\caption{As in Fig.\ref{fig:maps}, but for our smallest volume of 54 $\rm Mpc/h$ simulated 
with  $1024^3$, CUR3\_1024\_1.}
\label{fig:maps2}
\end{figure*}

\subsection{The simulated dataset}
\label{subsec:dataset}

Using different combinations of the physical modules described above, we explored the physical and numerical effects leading to the production and spreading of CRs in large-scale structures, from $z=30$ to $z=0$.  We simulated three independent cosmological volumes (with the side of $216 ~ \rm Mpc/h$,  $108 ~\rm Mpc/h$ and $54 ~\rm Mpc/h$, respectively) and for each of these we produced re-simulations with a different number of cells and DM particles. For the largest
investigated volume we used $2048^3$, $1024^3$ and $512^3$ cells, for the medium volume we used only a $1024^3$ cells and for the 
smallest volume we used $1024^3$, $512^3$ and $256^3$ grid cells. In every simulation we performed, the number of DM particles was the same as the number of grid cells. For a list of the various combinations of resolutions and box sizes, we refer the reader to Table 1. \\
In general, the effects of non-gravitational physics, such as radiative gas cooling and AGN feedback, are much more evident at high-resolution, where radiative cooling from free-free emission becomes comparable or shorter than the Hubble time given its scaling with $\propto \rho ^2$, and our scheme for AGN feedback begins to matter.
For this reason,  our study of the impact of non-gravitational physics is based largerly  
on the smallest box (with side $54 ~\rm Mpc/h$). On the other hand, the impact of different CR injection efficiencies on the outer cluster profiles is also captured at a lower
spatial resolution, and therefore all our boxes include a study using different
acceleration recipes. 
The use of such a composite set of volumes and resolutions is twofold: first, it allowed us to test the effect of several physical prescriptions on a small volume with the same spatial resolution that we also have in the largest volume ($308 ~\rm Mpc$) simulated with $2048^3$ cells {\footnote{3D renderings of a fly through the $2048^3$ run can be found at  the following URL: http://www.youtube.com/user/cosmofra}}, thus enabling a cheaper exploration of models and their application to the larger boxes after thorough testing. Second, convergence tests (see the following Sections) suggest that each box at its best resolution can be used for focusing on
slightly different topics: our largest box with a length of $308 ~\rm Mpc$ is useful to study outer cluster
profiles for a large number of objects ($\sim 10^2$); the medium-sized box measuring $154 ~\rm Mpc$ a side can be used to study cluster scaling relations and non-thermal emission from
galaxy clusters \citep[as in ][]{va13feedback}; the smallest box with a side of $77 ~\rm Mpc$ is more useful to study the turbulent outcome of AGN feedback models, and the impact of CRs and AGN feedback
on the cluster cores.

%

In summary, our simulations are designed to assess:

\begin{itemize}
\item the impact of different acceleration efficiencies of CRs at shocks, by comparing different choices of the function $\eta(M)$;
\item the impact of radiative cooling on the CR/gas energy budget in different cosmic environments;
\item the impact of AGN feedback on the energy of thermal gas and CRs in different cosmic environments;
\item the role of high-redshift SNe in the enrichment of CRs within galaxy clusters;
\item the impact of shock re-acceleration in comparison to direct shock injection;
\item the relative importance of weak ($M \leq 5$) shocks on the global enrichment of CRs;
\item the role of numerical spatial resolution in the final budget of CRs in large-scale structures.
\end{itemize}

Most of our simulations have been run on the {\small CURIE} BullX-Cluster hosted by GENCI in France, using the Thin Nodes architecture  {\footnote{http://www-hpc.cea.fr/en/complexe/tgcc-curie.htm}}.
Our largest box of $2048^3$ was completed on the Cray XC30 Piz Daint, at CSCS-ETH {\footnote{www.cscs.ch}} in Lugano (Switzerland).\\
Our runs took on average: a) $\sim 1500$ CPU hours for the $512^3$ boxes on $512$ processors; b) $\sim 20,000$ CPU hours for the $1024^3$ boxes on $1024$ processors; c) $\sim 1,200,000$ CPU hours for the $2048^3$ box on $4096$ processors. The exact duration of each run varied by factors of $\sim 2$ because of the inclusion of non-gravitational physics and/or the volume considered (i.e. due to the larger degree of non-linearity attained in the $54 ~\rm Mpc/h$ box, the final number of time steps was $\sim 4$ times larger than in the $216 ~\rm Mpc/h$ using the same number of cells).

\begin{table*}
\label{tab:tab2}
\caption{List of the simulations run for this project. Column 1: size of the simulated volume. Column 2: number of grid cells. Column 3: spatial resolution. Column 4: physical implementations used. Column 5: name of each run.}
\centering \tabcolsep 5pt 
\begin{tabular}{c|c|c|c|c}
 $L_{\rm box}$ [Mpc/h]& $N_{\rm grid}$ & $\Delta x$[kpc/h] & physics & ID \\  \hline
216 & $2048^3$ & $105$ & non-rad.+CR(KR13) & CUR1\_2048\_1 \\
 216 & $1024^3$ & $210$ & non-rad.+CR(KR13)& CUR1\_1024\_1\\
 216 & $1024^3$ & $210$ & non-rad.+CR(KJ07)& CUR1\_1024\_0\\
 216 & $1024^3$ & $210$ &  cool.+CR(KR13)+AGN & CUR1\_1024\_c1\\
 216 & $512^3$ & $420$ &  non-rad.+CR(KJ07)& CUR1\_512\_0\\
 108 & $1024^3$ & $105$ & non-rad.+CR(KR13) & CUR2\_1024\_1 \\
 108 & $1024^3$ & $105$ & non-rad.+CR(KJ07) & CUR2\_1024\_0 \\
 108 & $1024^3$ & $105$ & cool.+CR(KR13) & CUR2\_1024\_c1 \\
  54 & $1024^3$ & $52$ &  non-rad.+CR(KR13) & CUR3\_1024\_1)\\
  54 & $1024^3$ & $52$ &  non-rad.+CR(KJ07) & CUR3\_1024\_0\\
  54 & $1024^3$ & $52$ &  cool.+CR(KR13) & CUR3\_1024\_c1\\
  54 & $512^3$ & $105$ & non-rad.+CR(KR13) & CUR3\_512\_1 \\
  54 & $512^3$ & $105$ & non-rad.+CR(KR13) & CUR3\_512\_1 \\
  54 & $512^3$ & $105$ & non-rad.+CR(KJ07) & CUR3\_512\_0 \\
  54 & $512^3$ & $105$ & cool.+CR(KR13)+AGN & CUR3\_512\_c1 \\ 
  54 & $512^3$ & $105$ & cool.+CR(KJ07)+AGN & CUR3\_512\_c0 \\
  54 & $512^3$ & $105$ & cool.+CR(KJ07)+AGN+SN & CUR3\_512\_cSN0 \\
  54 & $512^3$ & $105$ & cool.+CR(KR13) & CUR3\_512\_c \\
  54 & $512^3$ & $105$ & non-rad.+CR(KR13),no reaccel.& CUR3\_512\_noreacc \\
  54 & $512^3$ & $105$ & non-rad.+CR(KR13) for $M>5$ & CUR3\_512\_1M5 \\
  54 & $512^3$ & $105$ & non-rad.+CR(KR13) for $M>3$ & CUR3\_512\_1M3 \\
  54 & $256^3$ & $210$ & non-rad.+CR(KR13)& CUR3\_256\_1 \\

\end{tabular}
\end{table*}

\section{Results}
\label{sec:res}

\subsection{Gravitational effects in non-radiative runs}
\label{subsec:non-rad}

Non-radiative simulations are our baseline model for the cosmological distribution of 
thermal and non-thermal energy. In this case, shock waves are the most important source of gas thermalisation and CR energy during structure formation, while the gravitational contraction of halos causes an additional increase of thermal and CR energy via adiabatic compression. \\

The large-scale distribution of gas and CR energy is shown in Figs.\ref{fig:maps}-\ref{fig:maps2}, where we present slices with a thickness of $L/8$ along the line of sight (where 
$L$ is the physical side of each box) for representative parts of the full volumes at $z=0$. 
These figures show that the cosmic rays are concentrated in the large-scale structure.
In this case the efficiency of \citet{kr13} has been adopted. 
While the thermal energy has a much larger volume filling factor, the CR energy is concentrated within large-scale structures, where it basically follows the thermal gas distribution, but  avoids the most rarefied regions because it is injected for the first time by accretion shocks.
These images also give a visual impression of the richness of clusters, groups and filaments reproduced in our largest box. Extracting virialised structures with a spherical halo-finder, we obtain catalogs of  $\sim 200$, $\sim 30$ and $6$ galaxy clusters with $M>10^{14} \rm M_{\odot}$ in the largest, in the middle and in the smaller box, respectively.\\

\bigskip

Cosmic rays can affect the distribution of the thermal gas in several ways.
First, by changing the local effective adiabatic index of the gas+CR mixture and 
the total pressure ($P_{\rm eff}=P_{\rm g}+P_{\rm cr}=\rho [(\gamma-1)e_{\rm g}+(\gamma_{\rm cr}-1)e_{\rm cr}]$. Second, the release of CRs at shocks reduces the thermalisation of the gas in the downstream region. Third, CRs lose their energies on cosmological timescales, via
hadronic and Coulomb losses, releasing a significant fraction ($\sim 1/6$) of their energy to the surrounding thermal gas.

\bigskip

The volume distributions of gas density and gas temperature are very similar, with the obvious trends related to the spatial resolution and the volume of each run (not shown): with increasing resolution (as in the ``CUR3'' case, where the resolution is $52 ~\rm kpc/h$) higher
densities (corresponding to the core of galaxy clusters) are reached, as in \citet{scienzo}. Within larger computational boxes higher temperatures are found, since larger volumes contain more massive clusters with higher virial temperatures. The temperature distributions  are insensitive to the adoption of a CR-injection efficiency. This suggests that for the bulk of the simulated cosmological volume the dynamical feedback of either of the two models is similar, and
very small (i.e. a few percent on the differential distributions).\\

\begin{figure*}
\includegraphics[width=0.97\textwidth]{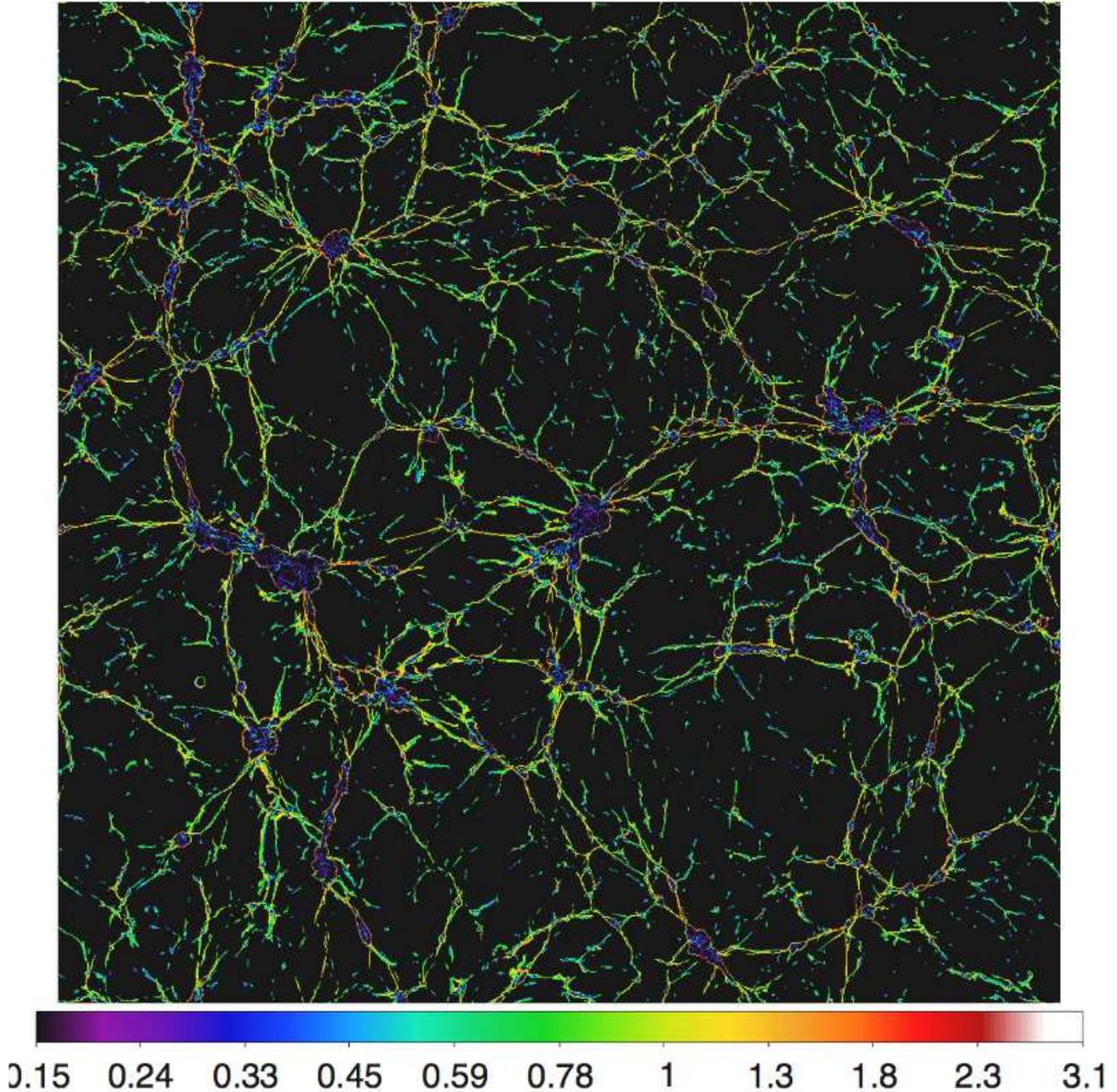}
\caption{Maps of shock Mach number (bottom, in $\log_{\rm 10} M$) for our $(216 ~\rm Mpc/h)^3$ simulated with $2048^3$ cells (CUR1\_2048\_1 at $z=0$). The image is $216 \times 216 ~\rm Mpc/h$ and is taken from a line-of-sight of 1 cell ($\approx 105 ~\rm kpc/h$).}
\label{fig:2048_mach}
\end{figure*}

Fig.\ref{fig:2048_mach} shows the reconstructed distribution of Mach numbers for a
slice through the $2048^3$ box of the CUR1 volume, with a thickness of 1 cell ($105~ \rm kpc/h$) along the line-of-sight.
Consistent with what has been reported in the literature \citep[e.g.][and references therein]{va11comparison}, our runs feature strong ($M>10-100$) accretion shocks outside of galaxy clusters, groups and filaments, and weaker ($M<10$) internal shocks connected to mergers and matter accretion within the virial volume of structures. This figure stresses the benefit of the use of large unigrid simulations which ensures that shocks are resolved everywhere at the best available resolution.

\begin{figure}
\includegraphics[width=0.45\textwidth]{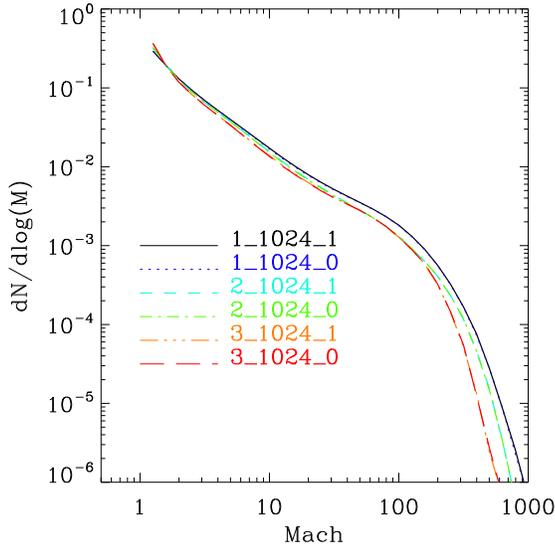}
\caption{Distribution of shocks with Mach number for our non-radiative $1024^3$ runs  at $z=0$. The CUR1\_1024\_1 and CUR1\_1024\_0 runs are for a $(216 ~\rm Mpc/h)^3$ volume, the CUR2\_1024\_1 and CUR2\_1024\_0 runs are for a $(108 ~\rm Mpc/h)^3$ volume and the CUR3\_1024\_1 and CUR3\_1024\_0 runs are for a $(54 ~\rm Mpc/h)^3$ volume.}
\label{fig:mach1}
\end{figure}

The spatial distribution of Mach numbers for all $1024^3$ runs of our project
are compared in  Fig. \ref{fig:mach1}.  
The distribution of Mach numbers follows a steep power-law 
with $\alpha \sim -1-1.5$ 
($\alpha = d\log N(M)/d\log M$) for $1 \leq M \leq 100$, which considerably steepens for stronger shocks. 
With increasing spatial resolution, the median Mach number moves to lower values and the Mach number distribution becomes steeper (see also  \cite{va11comparison}). The reason for this is that the Mach number estimate is based on changes in hydrodynamic variables over a certain length scale that depends on the resolution of the simulation. For example, the gas density increases as the accreting matter flows onto a galaxy cluster or a filament. First, it gets adiabatically compressed and then increases discontinuously as it goes through the shock transition. With coarser grid cells, one averages over larger spatial scales. Hence, at low spatial resolution the average density estimated for pre-shock region will also include some of the isentropic increase in the density in the upstream region, leading to an overestimate of the Mach number.

\begin{figure}
\includegraphics[width=0.45\textwidth]{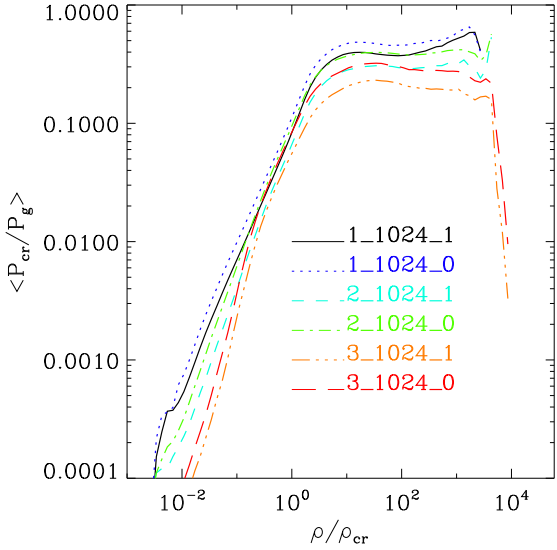}
\caption{$X_{\rm cr}$ as a function of gas overdensity for the same runs shown in Fig.\ref{fig:mach1}, where the results for the $(216 ~\rm Mpc/h)^3$, $(108 ~\rm Mpc/h)^3$ and $(54 ~\rm Mpc/h)^3$ are compared.}
\label{fig:distr_cr1}
\end{figure}

The average pressure ratio $X_{\rm cr}=\langle P_{\rm cr}/P_{\rm g} \rangle${\footnote{This estimate is used instead of $\langle P_{\rm cr}\rangle/\langle P_{\rm g} \rangle$, that in runs  with radiative cooling can be biased by the presence of a few massive objects with cooling flow not fully quenched by AGN feedback, while the adopted estimate instead is suitable to describe the average properties of the full cosmic range of densities.}} as a function of gas density is shown in Fig.\ref{fig:distr_cr1}. \\
Within large-scale structures ($\rho \geq 1-10 ~\rho_{\rm cr}$) the pressure support
from CRs injected by cosmological shocks ranges from $\sim 10$ to $\sim 50$ percent of the thermal pressure, 
depending on the resolution and acceleration efficiency.
The lowest pressure support is always found in the case of the \citet{kr13} model. 
$X_{\rm cr}$ shows a significant change with resolution/volume especially at high density and in the very rarefied
cosmic environment. As we will see also in Sec.\ref{subsec:resol} the first trend is mainly driven by the better
resolution of cluster cores, while the second is due to the reduced strength of outer accretion shocks discussed above. Across several orders of magnitude
in gas density, the pressure support from CRs is reduced by a factor of $\sim 3-4$ if the
spatial resolution is increased by a factor $4$. In our most resolved runs (``CUR3\_1024\_0'' and ``CUR3\_1024\_1'') the pressure support from CRs is small within
the innermost cluster regions, $X_{\rm cr} \leq 10$ percent inside $R_{\rm 500}$ and $X_{\rm cr} \leq 1$ percent within $R_{\rm 2500}$, as can be seen from the drop of $X_{\rm cr}$ in very high density range in Fig.\ref{fig:distr_cr1}{\footnote{$R_{\rm 500}$ and $R_{\rm 2500}$ are defined as the radius within which the average density of the cluster is $500$ and $2500$ than the cosmological critical density.}}. In this case, the maximum support
from CRs is found at about $\sim 10 \rho_{\rm cr}$, where $X_{\rm cr} \sim 0.3$ in the
\citet{kj07} model and $X_{\rm cr} \sim 0.2$ in the \citet{kr13} model.
The pressure support from CRs shows little evolution with spatial resolution, and the range $0.1 \leq \rho/\rho_{\rm cr} \leq 5$.

In the case of non-radiative simulations, the drop of the CR to gas pressure in the centre of galaxy clusters
is due to the fact that most of the shock energy dissipation in clusters is mediated by weak shocks, $2 \le M \le 3$ \citep{ry03,pf06}, for which the injection efficiency is rather small in the DSA models investigated here. Typically,
an increase in resolution leads to a better reconstruction of weak shocks \citep[e.g.][]{va11comparison}, and to 
a decrease in the CR energy dissipation \citep[e.g.][]{va11comparison}.
The rather abrupt drop in the average value of $X_{\rm cr}$ for the resolution of $52~ \rm kpc/h$ suggests that in order to compute the energy budget of CRs cluster cores accurately a better resolution is necessary. A more detailed examination of the cluster core properties will be the subject of a forthcoming paper using the same set of simulations.

\begin{figure}
\includegraphics[width=0.45\textwidth]{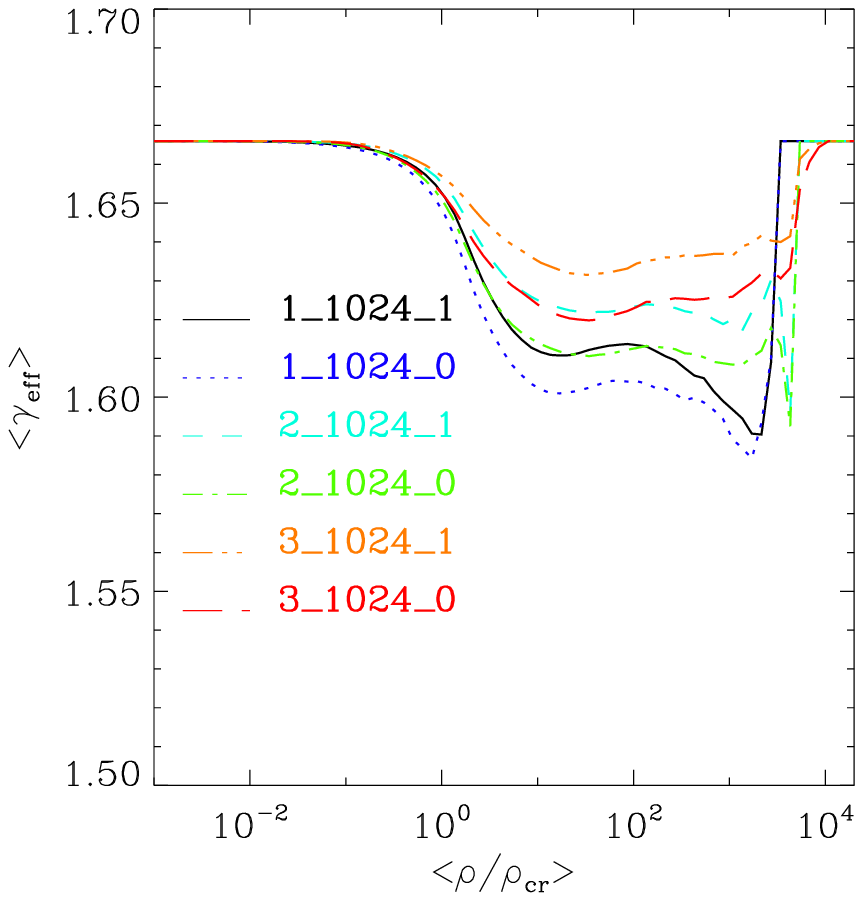}
\caption{$\langle \gamma_{\rm eff}\rangle$ as a function of gas overdensity for the same runs shown in Fig.\ref{fig:mach1}, where the results for the $(216 ~\rm Mpc/h)^3$, $(108 ~\rm Mpc/h)^3$ and $(54 ~\rm Mpc/h)^3$ are compared.}
\label{fig:distr_gamma1}
\end{figure}

Different levels of $X_{\rm cr}$ produce a different degree of modification in the
local effective adiabatic index, $\langle \gamma_{\rm eff}\rangle$ (Sec.\ref{subsec:cr}). Fig.\ref{fig:distr_gamma1} shows the trend with gas density of the effective adiabatic
index for these runs. The departures from the non-relativistic value $5/3$ are never large and decrease with increasing resolution. In our most resolved run, the average index is $\sim 1.63$ 
within large-scale structures in the \citet{kj07} models and $\sim 1.64$ in the \citet{kr13} model.

\subsection{Additional non-gravitational effects}
\label{subsec:cool}

The impact of radiative cooling, AGN and SNe are studied through a larger set of re-simulations of the same
volumes presented in the previous Section. 
We first consider the runs where extra physics acts on the same spatial resolution ($104 ~ \rm kpc/h$) within the same CUR3 volume. Then, we also consider runs in which the prescriptions for the injection of CRs
are modified. \\

\subsubsection{Runs with radiative cooling and feedback from AGN and SNe}

\begin{figure*}
\includegraphics[width=0.97\textwidth,height=0.45\textheight]{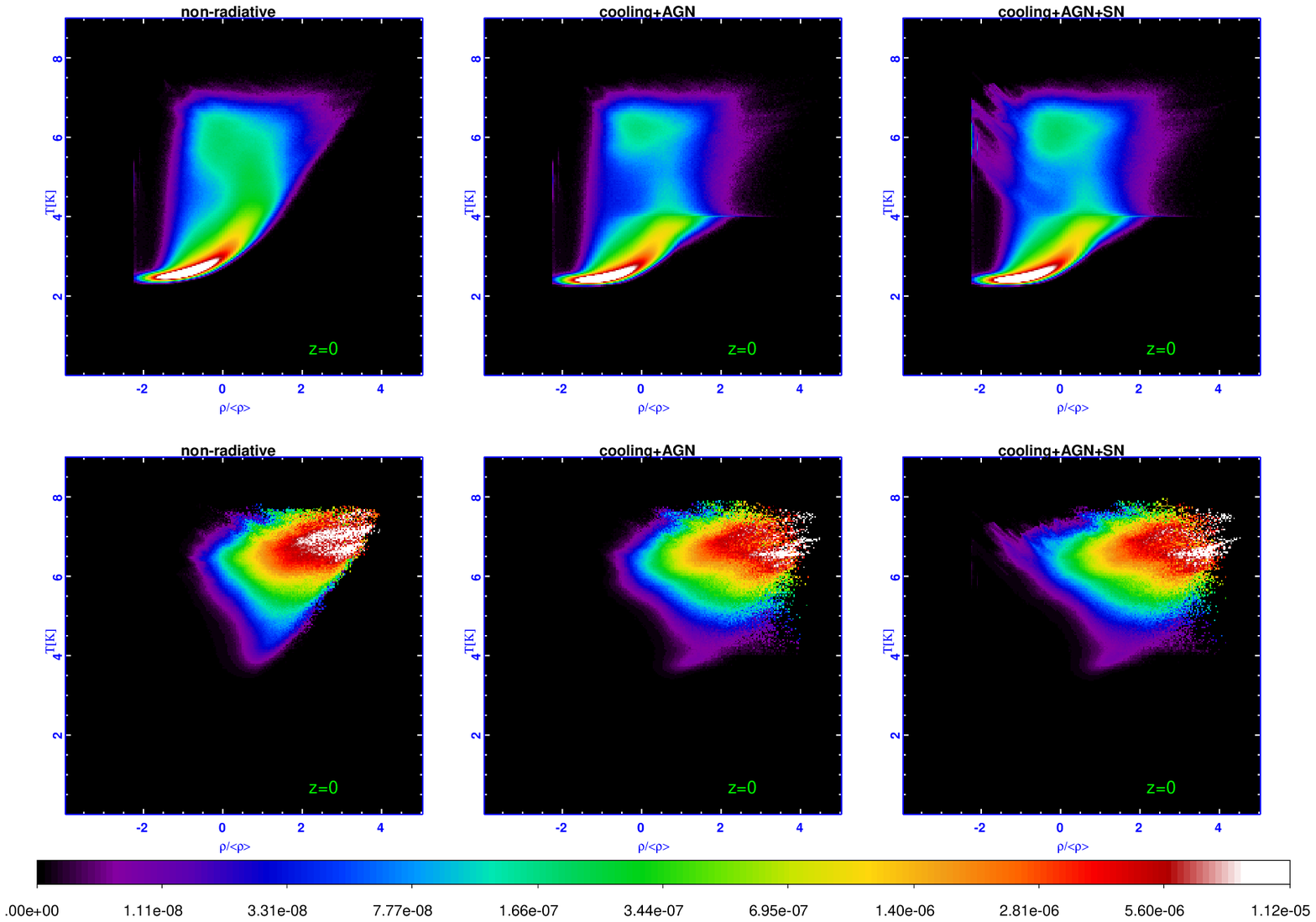}
\includegraphics[width=0.97\textwidth,height=0.45\textheight]{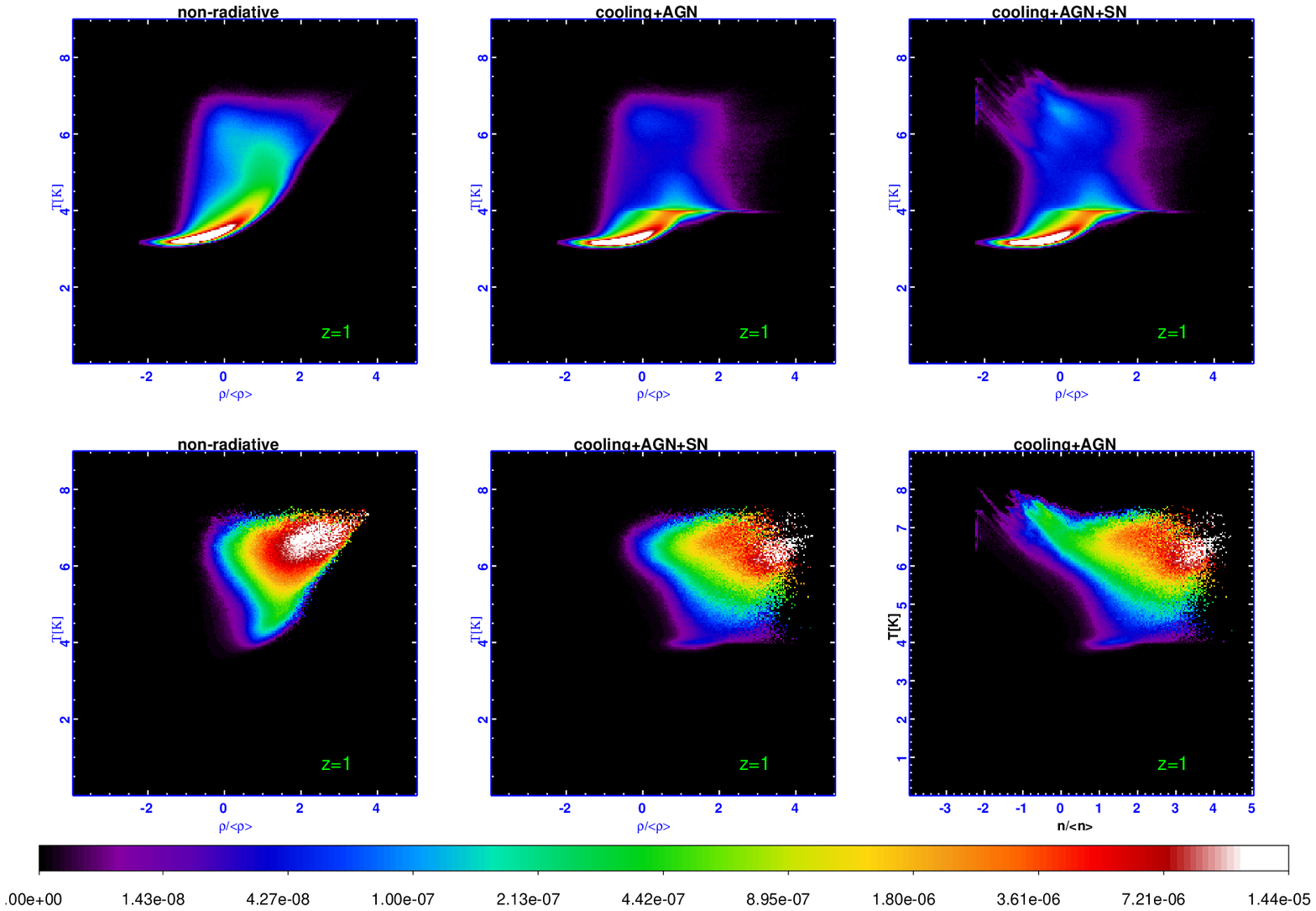}
\caption{Phase diagrams for the CUR3 run with $512^3$ cells at $z=0$, for the non-radiative re-simulation (left), for the cooling+AGN re-simulation (centre) and for the cooling+AGN+SN re-simulation (right).
In the top panels the volume within each cell is used as a weighting field, in the lower panels the CR energy within the cell is used instead. The lower set of panels show the same statistics for $z=1$.}
\label{fig:phases}
\end{figure*}

Cosmological runs employing radiative gas cooling and non-gravitational energy feedback from galactic activity and star formation still represent a complex challenge because
it is presently impossible to fulfill all observational constraints within the same
numerical model in cosmology \citep[e.g.][for a recent review]{2012ARA&A..50..353K} \\
In this study we followed the simplistic approach of including the dynamical effect of feedback from AGN (identified as the highest density peaks inside clusters, Sec.\ref{subsubsec:agn}) and from a population of SNe at $z=2$ (Sec.\ref{subsubsec:sn}).
These additional processes first modify the density and temperature distribution within
the cosmological volume by enhancing the compression of gas and by increasing the amount of  the cold
gas phase (via radiative cooling). Secondly, they introduce  localised feedback events that release additional thermal/CR energy in large-scale structures from inside-out. In this Section we focus on the various re-simulations of the $(54 ~\rm Mpc/h)^3$ volume with $512^3$ cells, which allows a parameter study within a reasonable computing time.
A closer look at the phase diagrams ($T$ vs $\rho$) highlights some
important effects of non-gravitational physics that are less evident in the spatial distribution (Fig. \ref{fig:phases}). Here we compare the outcomes of the AGN feedback (central panels) and AGN plus SN feedback (right panels) in comparison to the baseline non-radiative run with CR-physics only (left panels) at $z=1$ and $z=0$.
The phase diagrams are weighted by the cell volume (top panels for each redshift) and by the CR energy within the cell (lower panels of each redshift). In both approaches we normalise the intensity to the total gas mass/CR energy within the volume.
The energy of CRs is mostly located within hot ($T>10^7 \rm K$) and overdense ($\geq 10^2 \langle \rho \rangle$) regions in non-radiative runs.
When the cumulative effect of a population of SNe at $z=2$ is considered,  gas and CRs start to get expelled from halos at $z=1$ (and to a lesser extent also at $z=0$), as a hot and underdense gas phase which is absent in the other runs.
The total energy carried by this phase is rather small ($\leq 10^{-4}-10^{-3}$ of the
thermal/CR energy within halos), yet its impact on the rarefied environment into which the outflow
is expanding can be rather strong since it drives strong expanding shocks and enhances the thermalisation outside of the ICM \citep[e.g.][]{ka07,va13feedback}.\\
The signature of high-redshift supernovae in the phase diagram is qualitatively in agreement with similar phase diagrams obtained through higher-resolution simulations of galaxy formation \citep[][]{2006ApJ...641..878T,2013MNRAS.432.1989S}.\\

\begin{figure}
\includegraphics[width=0.4\textwidth]{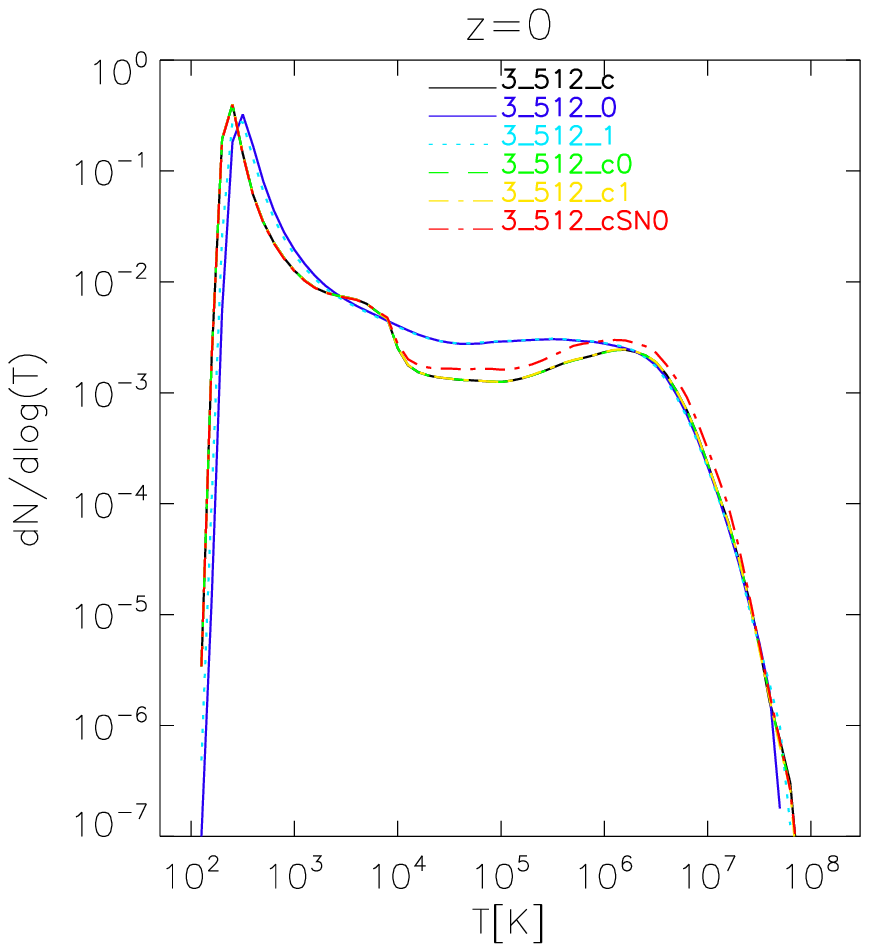}
\includegraphics[width=0.4\textwidth]{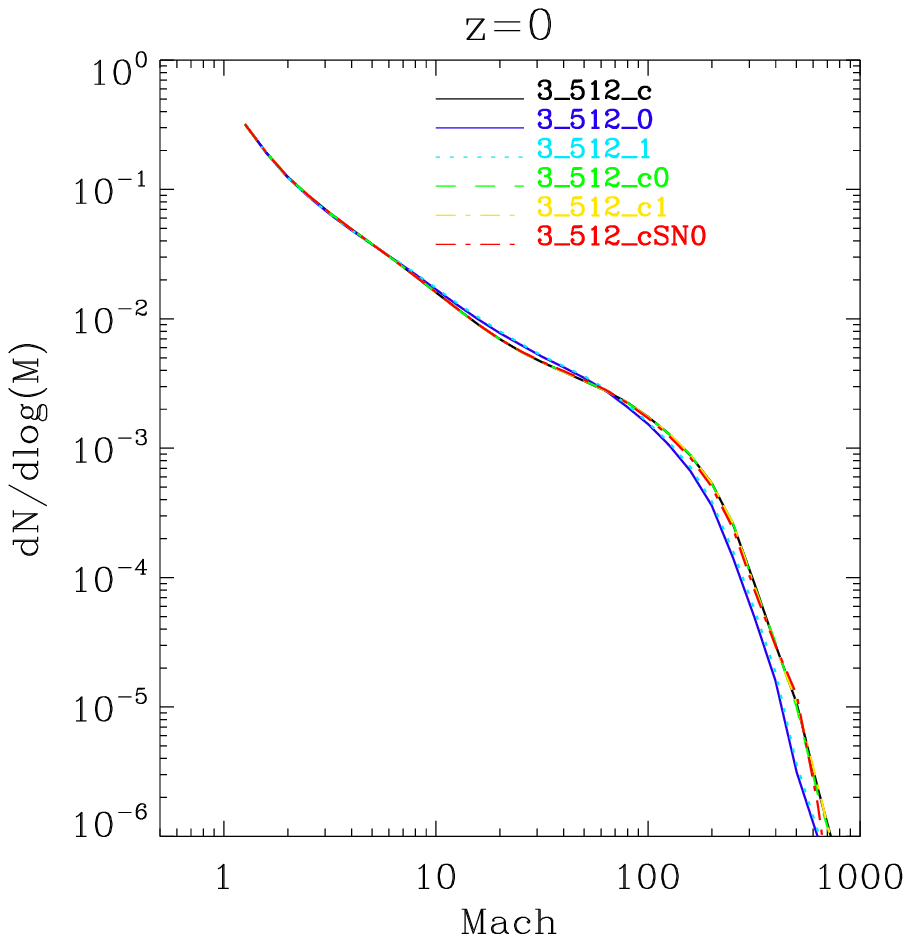}
\includegraphics[width=0.4\textwidth]{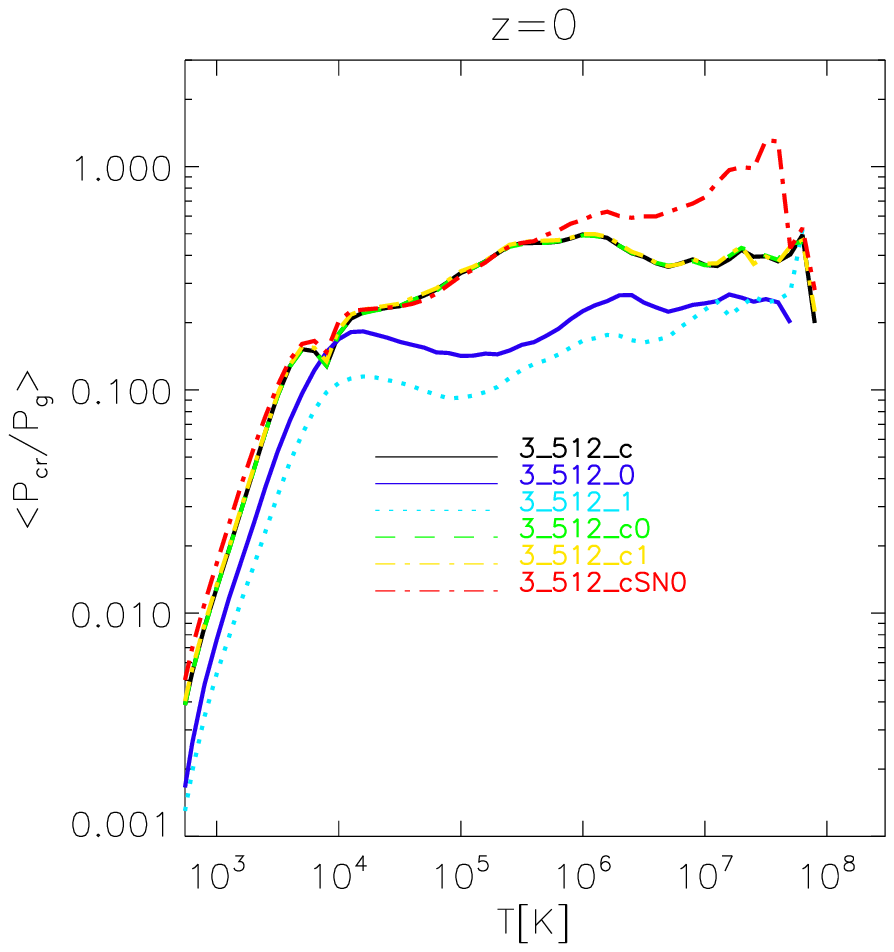}
\caption{Top and middle panels: volume distribution function of gas temperature and Mach numbers for the different runs of the $(54 ~\rm Mpc/h)^3$ at $z=0$ employing a box of $512^3$. Lower panels: $X_{\rm cr}$ as a function of gas temperature for the same runs.}
\label{fig:distr_cr2}
\end{figure}

The effect of baryonic physics on the temperature distribution of the various CUR3 runs with $512^3$ cells is
shown in the top panel of Fig.\ref{fig:distr_cr2}. The effect of radiative cooling with respect to the non-radiative run is evident in the range $10^4 \rm K \leq T \leq 10^7 \rm K$. We notice that SNe contribute to a small temperature excess in the same range, compared to runs including AGN only.
The second panel within the same figure shows the spatial distribution of Mach numbers for the same runs. Different physical implementations lead to very similar results, confirming that the majority of shocks in large-scale structures is primarily a by-product of gravitationally induced motions, and that radiative
cooling does not alter the above picture \citep[e.g.][]{pf06,va09shocks,sk08}. Feedback from AGN and SNe is only responsible for a small excess of strong external shocks with $M>10$. This is mostly caused by winds that non-gravitational effects produce when supplying the ICM with an excess of thermal energy. A similar effect (even if on a quantitatively different level, due to a difference in the feedback scheme) has been reported by \citet{ka07}, and also in our earlier AMR simulations \citep[][]{va13feedback}.\\
The third panel of Fig.\ref{fig:distr_cr2} shows the average pressure ratio of CRs as a function of temperature
for the same runs. SNe at $z=2$ do not alter significantly the thermal and CR energy
budgets in large-scale structure, but have a significant impact on the energy
budget of  the inter galactic medium (IGM) outside of virialized halos. This appears as a peak of $X_{\rm cr}$ at $\sim 3-5 \cdot 10^7 ~\rm K$ in the phase diagrams.
If SNe are not included, the average trend of $X_{\rm cr}$ is very similar if pure cooling runs or runs with AGN are compared. This suggests that a significant increase of $X_{\rm cr}$ with respect to the
non-radiative runs is mainly due to the loss of thermal energy because of radiative cooling, and not to AGN
effects.
However, The adoption of AMR is expected to increase this difference,  since at a higher resolution the AGN feedback is triggered more frequently and earlier in time \citep[][]{va13feedback}.\\

In reality, the effect of CR diffusion (which we do not model here) is likely to increase the density of CRs in the outer IGM, provided that the magnetic field carried by the outflows is small enough to allow for the diffusive escape of particles.
A recent study by \citet{beck13} suggested that up to $\sim 10^{-6} \mu G$ of magnetic field can be carried into the voids by star-forming galaxies. However, this estimate can vary by a few orders of magnitude depending on where the winds start \citep[][]{2006MNRAS.370..319B}.
The CRs are also expected to generate additional magnetic fields by driving electric currents, at a rate of $\sim 10^{-17} \rm G/Gyr$ \citep[][]{mb11}.  A lower limit of $\sim 10^{-7} \mu G$ has been derived from the non-observation of $\gamma$-ray emission from electromagnetic cascade initiated by tera-electron volt gamma-rays in IGM by \citet{2010Sci...328...73N}. Such magnetic fields are weak enough to allow for a fast diffusion of CRs on $\gg \rm Mpc$ scales outside of clusters and filaments. \\

In summary, within the several physical models explored in this work, the one including an early contribution from
SNe is responsible for the largest energy budget in CRs in the low-density Universe, especially at high redshift. On the other hand, at low redshifts, the pressure ratio of CRs within large-scale structures
is primarily affected by radiative cooling, which is always found to increase $X_{\rm cr}$ with respect to the
non-radiative case, by radiating away the thermal gas energy.
In general, the impact of the distributed modes of energy feedback from AGN and SNe on the pressure ratio of CRs is small at low redshift and limited to overdensities of $\rho \gg 10^3 \rho_{\rm cr} $.

\begin{figure}
\includegraphics[width=0.45\textwidth]{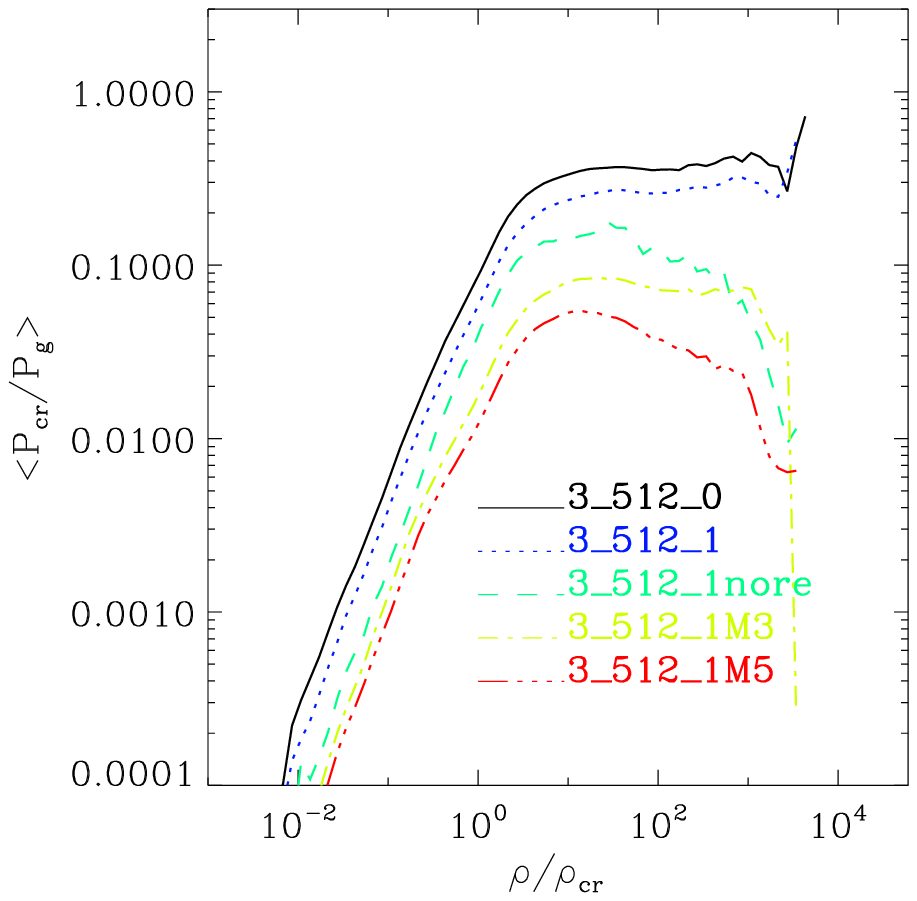}
\includegraphics[width=0.45\textwidth]{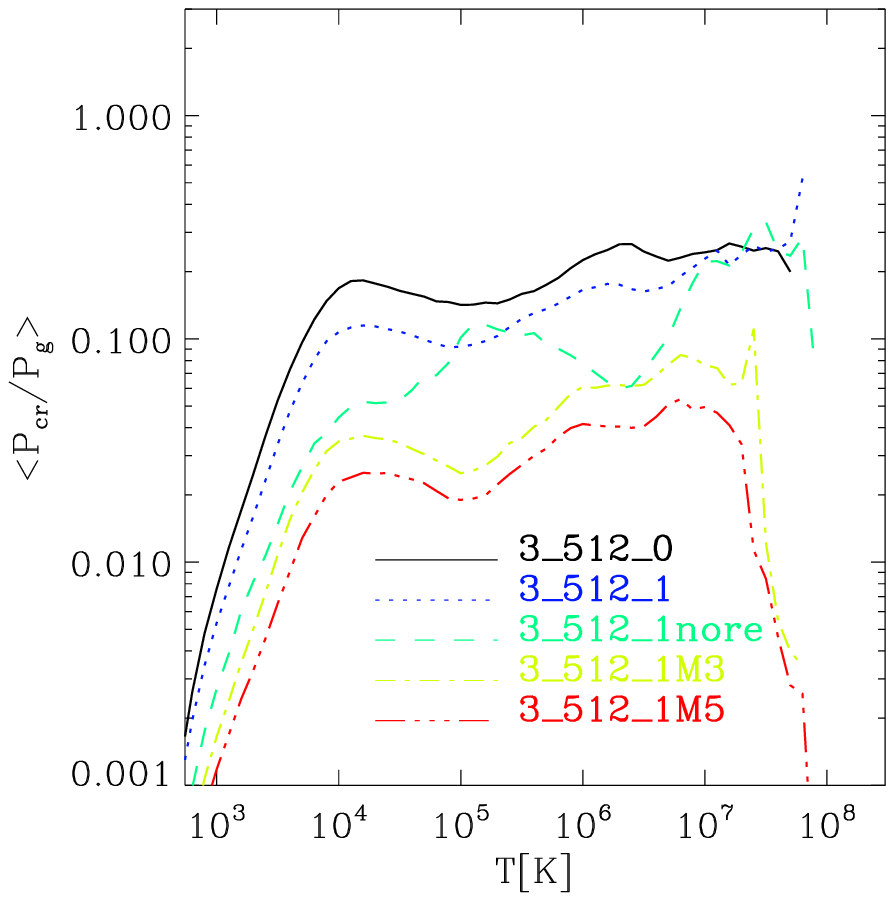}
\caption{$X_{\rm cr}$ as a function of gas overdensity (top panels) and gas temperature (bottom panels) for re-simulations of the  $(54 ~\rm Mpc/h)^3$ volume adopting different models of CR injection at low Mach numbers.}
\label{fig:distr_cr3}
\end{figure}

\subsubsection{Runs with different prescriptions for the injection of CRs}
\label{subsubsec:crs}

In a second set of $512^3$ runs of the $54 ~\rm Mpc/h$ volume we studied the effects of several prescriptions
of DSA  in the final budget of CR energy.
First, we test the effects of re-acceleration (with run ``CUR3\_512\_1nore'', where we used $\eta(M)$ for the $E_{\rm cr}/E_{\rm g}=0$ case) and of a higher threshold in Mach number
for the injection of CRs (runs ``CUR3\_512\_1M3'' and ``CUR3\_512\_1M5''), using the \citet{kr13} model as a reference. These runs probe the role of  $M \leq 5$ shocks in enriching the ICM with CRs. 
This is particularly relevant, since the bulk of shock thermalization in the cosmological volume
happens at these shocks \citep[][]{ry03,va11comparison}, and the injection of relativistic protons and electrons out of the
thermal pool is an open issue in theoretical models \citep[e.g.][for a recent discussion]{kr13,2013MNRAS.435.1061P,2013arXiv1307.4754V}, with large uncertainties in the total acceleration
efficiency.

Fig.\ref{fig:distr_cr3} presents the gas density and temperature distribution of $X_{\rm cr}$ for these 
runs. The decrease of CR injection at weak shocks steepens the average relation
$X_{\rm cr}(\rho/\rho_{\rm cr})$, starting from $\rho \geq 10 ~\rho_{\rm cr}$,
while it almost uniformly lowers the value  of $X_{\rm cr}$ at all temperatures. 
For the typical conditions in galaxy clusters, the pressure contribution from CRs is reduced by a factor of
$\sim 20-30$ if a minimum threshold of $M_{\rm thr}=5$ is adopted, and by $\sim 10$ if $M_{\rm thr}=3$ is
adopted. This confirms the predominant role of weak (mostly merger) shocks in the enrichment of CRs within the ICM.
At this spatial resolution re-acceleration is significant only
for $\rho \geq 10^2 ~\rho_{\rm cr}$, i.e. within the virial region of clusters. Outside large-scale structures ($\rho \leq 10~\rho_{\rm cr}$), the effect of these different assumptions is to shift 
the normalization of the average dependence of $X_{\rm cr}$. This suggests that  strong $M \gg 5$ shocks
are responsible for setting the average trend with gas density/temperature in the cosmic volume. Furthermore, the impact
of weak shocks is not negligible even in these rarefied environments. Indeed, while the cluster outskirts are
characterized by strong accretion shocks, the accretion shocks around filaments can be rather weak, owing
to their smaller overdensity and gravitational potential (e.g. Fig. \ref{fig:2048_mach}).\\
In summary, these runs bracket the still existing uncertainties in the acceleration efficiency
of CRs at cosmological shocks. The maximum pressure ratio of CRs is found 
for the \citet{kj07} model, $\sim 30-40$ percent inside large-scale structures at this resolution, while
the lowest pressure ratio is found for the \citet{kr13} model with no acceleration for $M \leq 5$, with
$\sim 1-5$ percent inside large-scale structures. Interestingly, the slope of the average trends of $X_{\rm cr}$ with gas density and temperature is unaffected by the explored changes in the acceleration
efficiency for $\rho < 10 \rho_{\rm cr}$ (e.g. $X_{\rm cr} \propto \rho $). For larger densities the
slope of the relation changes from being flat to a smooth decrease with density, if the impact
of weak shocks (and/or reacceleration) is changed. This latter density range is more relevant to galaxy
clusters, and suggests that in principle an observable related to $X_{\rm cr}$ (like hadronic $\gamma$-ray
emission) can be used to constrain the functional form of the acceleration efficiency $\eta(M)$. However,
at present only upper-limits of $\gamma$-ray emission from galaxy clusters have been reported \citep[][]{ack10,2013arXiv1308.6278H,fermi13}, and this information is mostly limited to the cluster core regions ($\rho>2500 \rho_{\rm cr}$).

\begin{figure*}
\includegraphics[width=0.33\textwidth]{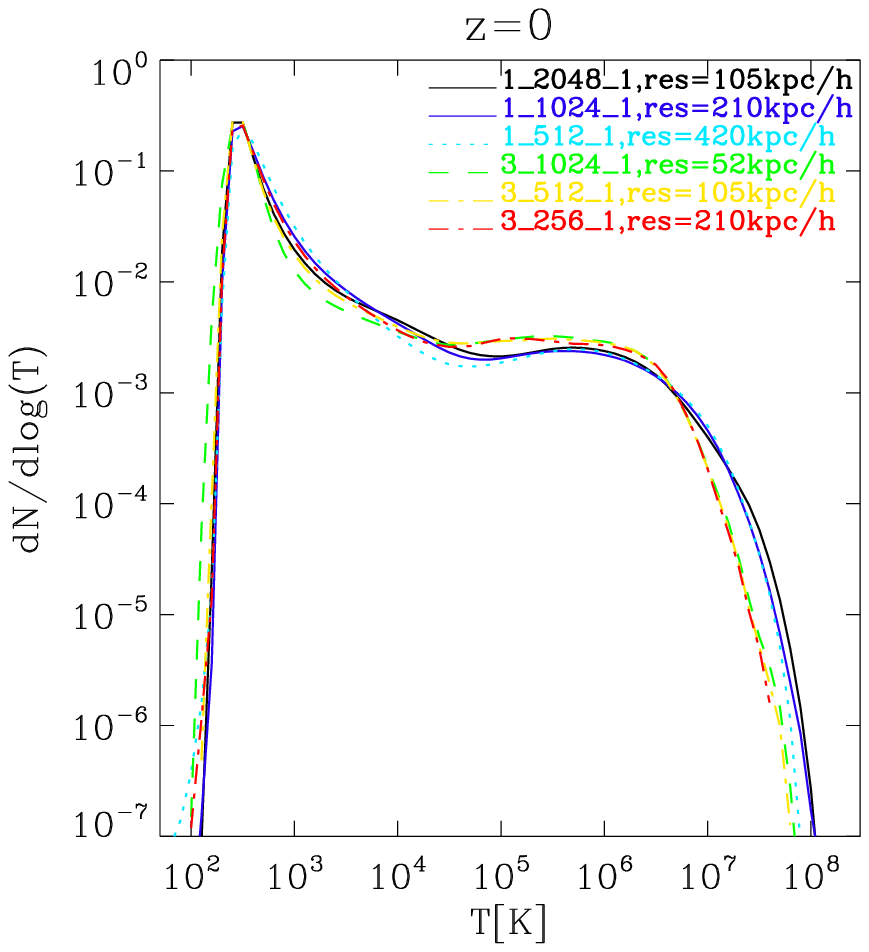}
\includegraphics[width=0.33\textwidth]{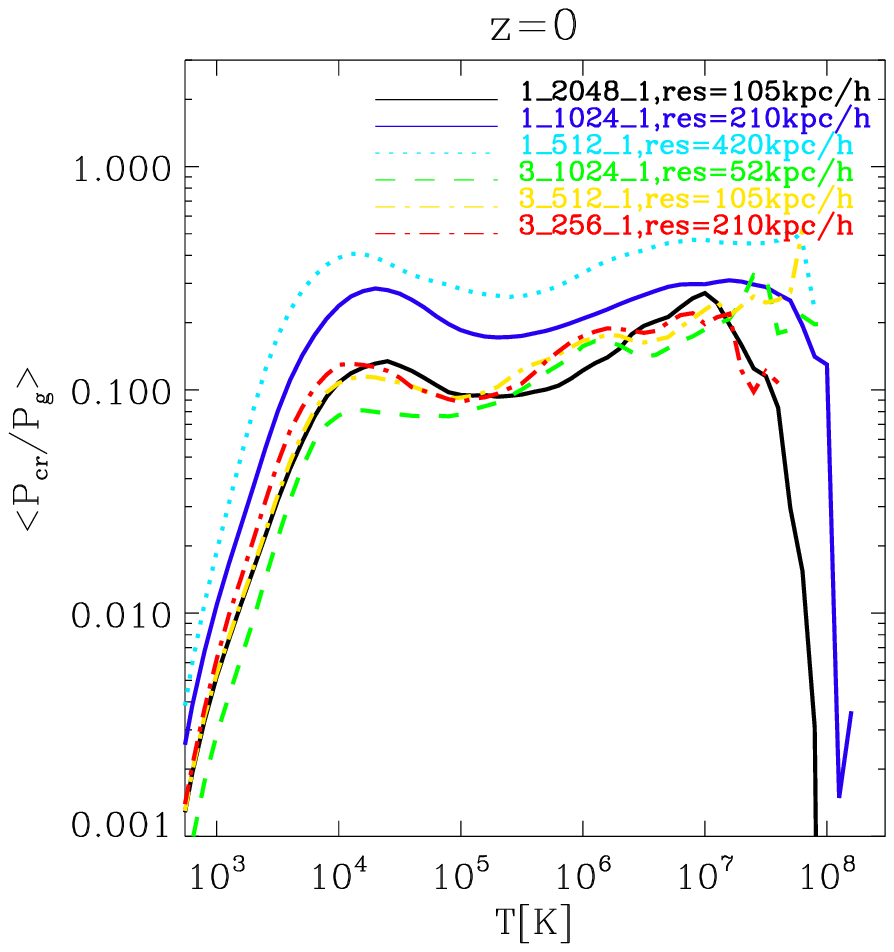}
\includegraphics[width=0.33\textwidth]{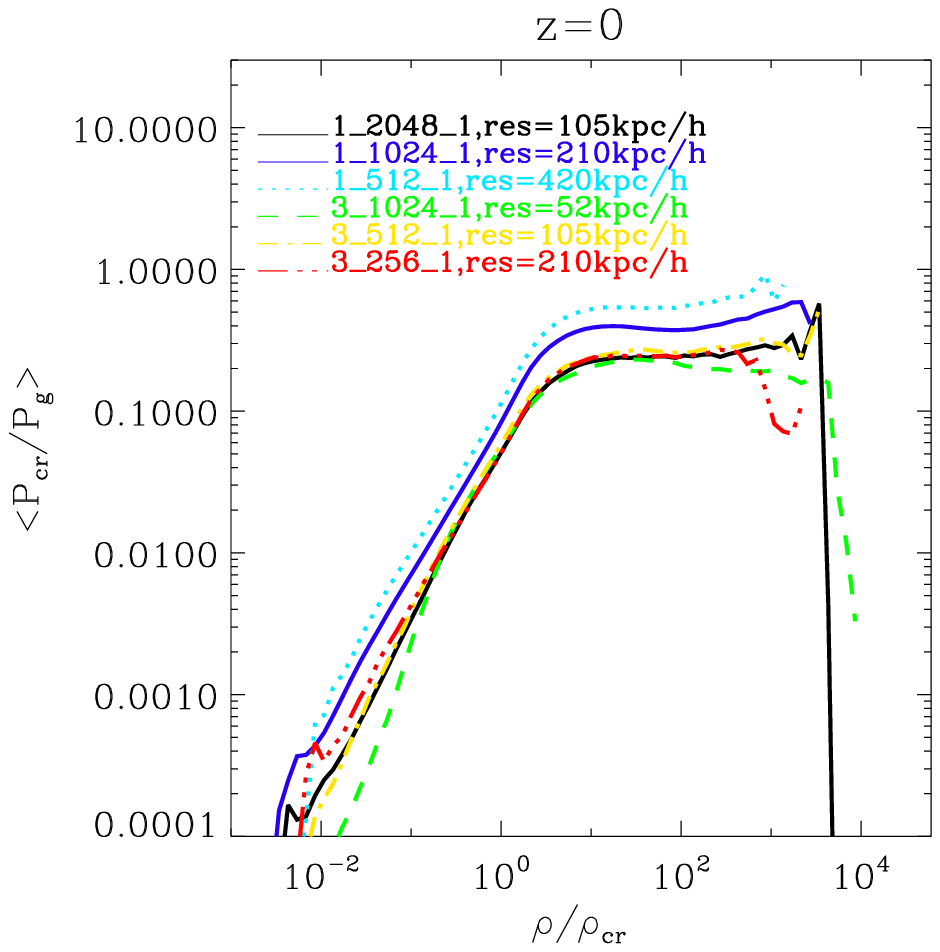}
\caption{Left panel: volume distribution (number of cells over total volume) of gas temperature for the CUR1 and CUR3 volumes at $z=0$, considering all available resolutions. Central and right panels: 
$X_{\rm cr}$ as a function of gas temperature (middle) and gas overdensity (lower) for the same runs. In all cases the acceleration efficiency of \citet{kr13} is assumed.}
\label{fig:distr_cr4}
\end{figure*}

\subsection{Resolution effects}
\label{subsec:resol}

Spatial resolution may change the average strength of shock waves in the ICM (Sec.\ref{subsec:non-rad}), and consequently the average efficiency of CR-injection in simulations.

Fig.\ref{fig:distr_cr4} shows a resolution study for the largest and the smallest volumes of our project, where
we probed the effect of spatial resolution from $\Delta x = 420 ~\rm kpc/h$ to $\Delta x = 52 ~\rm kpc/h$. The 
distributions of gas temperature show that the two simulations are rather converged already for a resolution of $\Delta x = 210 ~\rm kpc/h$. The distributions in the two volumes are very similar, except in the high-temperature
range, given the larger virial temperatures of the high-mass clusters in run CUR1. 
The average distribution of $X_{\rm cr}$, instead, shows hints of a slower convergence
with resolution.  
For most of the cluster volume ($\rho/\rho_{\rm cr} \leq 10^2$ and $T \leq 10^7 \rm K$)
$X_{\rm cr}$ is reasonably converged as soon as the uniform spatial resolution goes below $\Delta x \leq 105 ~\rm kpc/h$.  The further increase of resolution obtained in the CUR3 run produces a significant evolution of $X_{\rm cr}$ only for the overdensity typical of cluster cores. However, the trend in this box is partially affected by 
the small number statistics of massive galaxy clusters (i.e. $\sim 8$ with $M \geq 10^{14} M_{\odot}$), and dynamical differences between these few objects can amplify smaller differences in the trend. 

In summary, these results suggest that for a uniform spatial resolution of the order of $\sim 100 ~\rm kpc/h$, the 
average dynamical impact of CRs onto large-scale structures is converged, once the acceleration efficiency is fixed. At this resolution, the most energetic shock waves in the cosmic volume are properly resolved \citep[][]{ry03,va11comparison}, and the distribution of Mach numbers is reasonably converged. A further increase
in resolution causes a significant decrease in the pressure ratio of CRs only for $\rho \geq 10^3 \rho_{\rm cr}$, i.e. for the core regions of galaxy clusters. For such high resolution, however, also the impact of AGN feedback is expected to be significant in setting the level of CRs within cluster cores.
The observed trends are in agreement with our earlier study \citep[][]{scienzo}, that used a different acceleration efficiency but not reach such a high resolution.

\subsection{Cosmological Evolution}
\label{subsec:time}

\begin{figure*}
\includegraphics[width=0.45\textwidth]{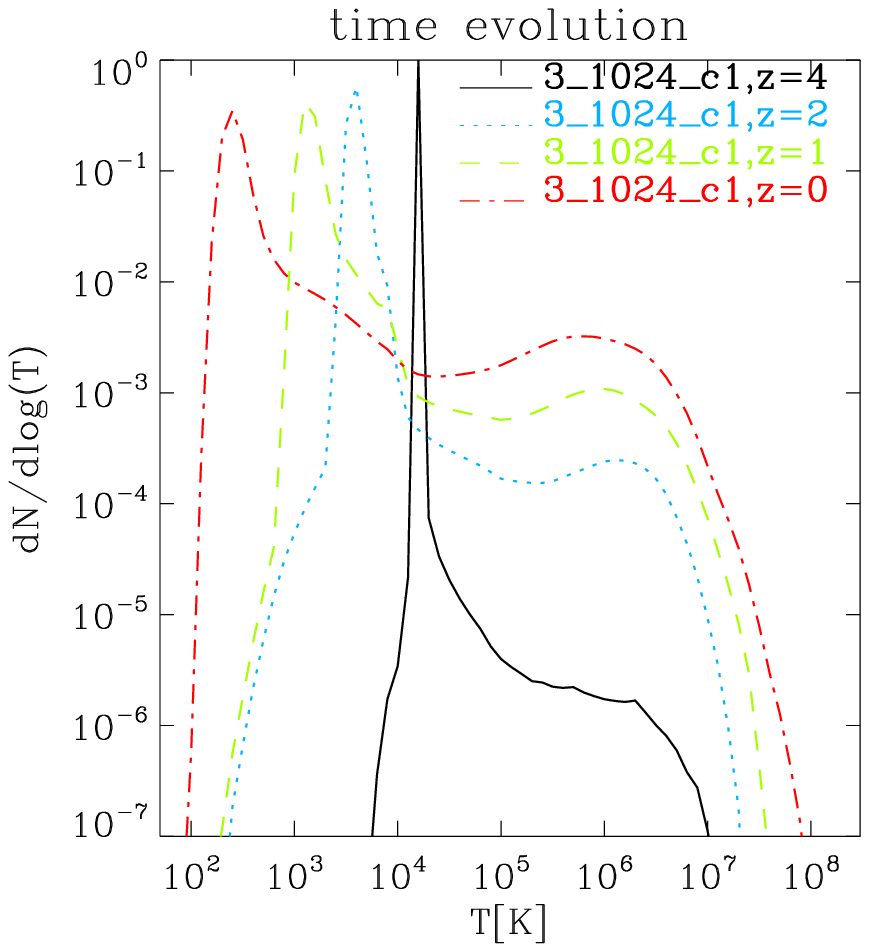}
\includegraphics[width=0.45\textwidth]{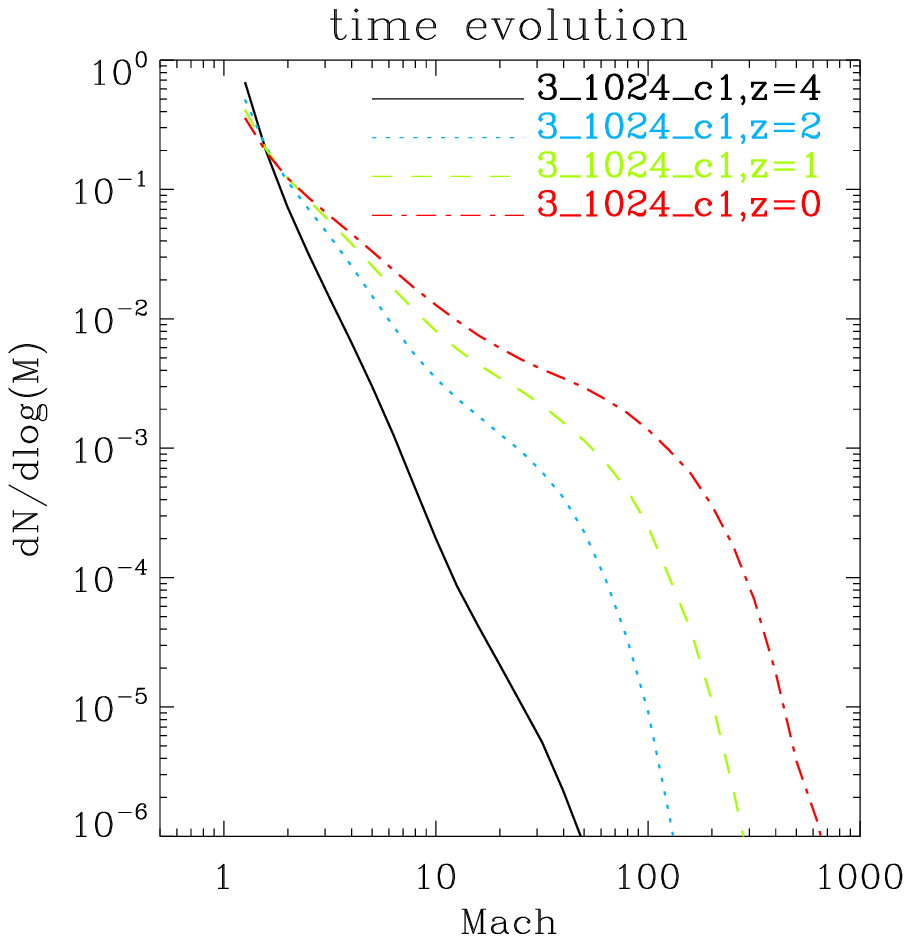}
\includegraphics[width=0.45\textwidth]{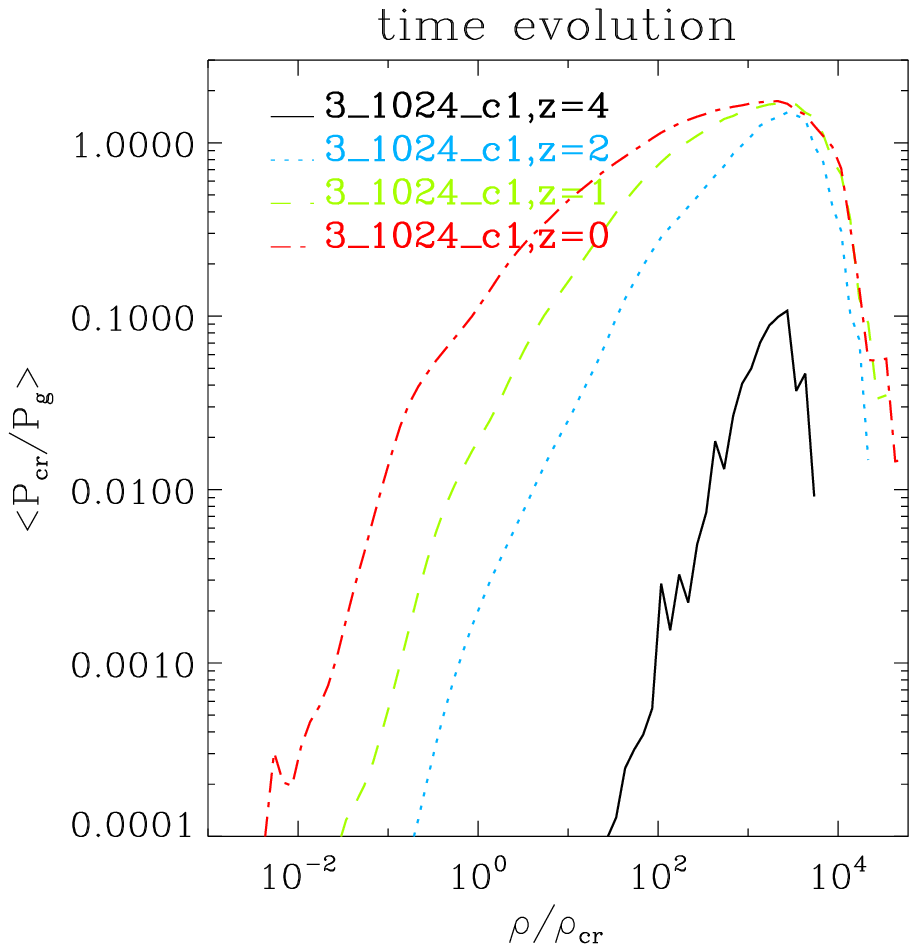}
\includegraphics[width=0.45\textwidth]{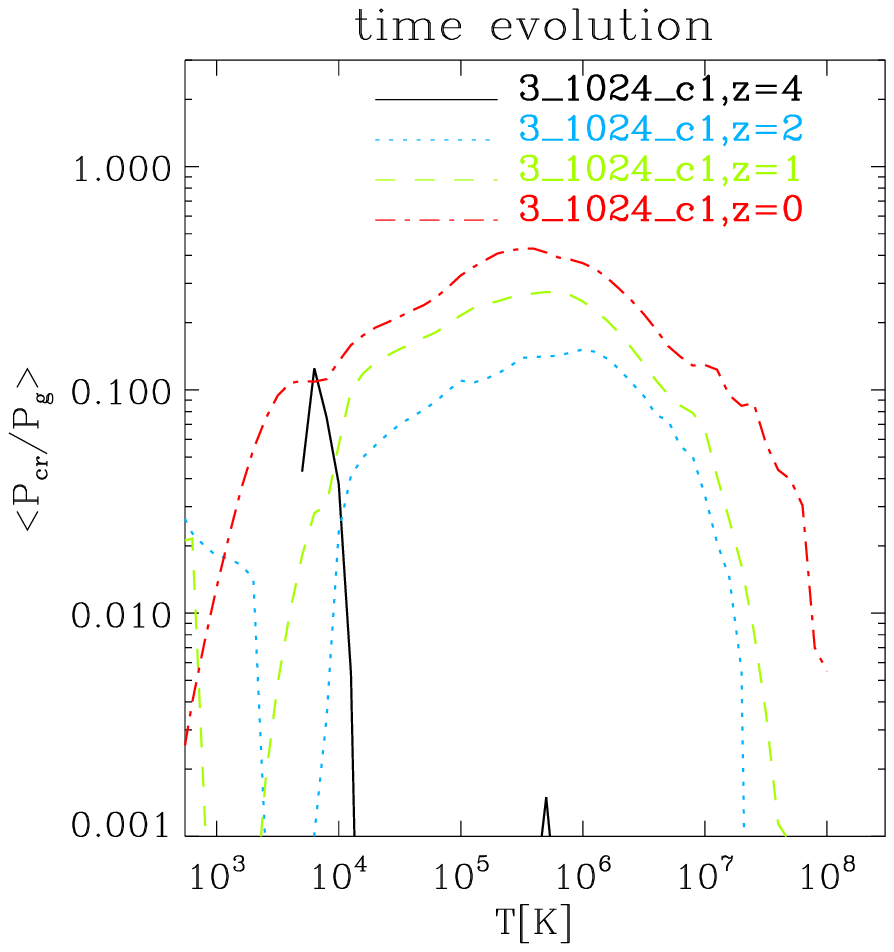}
\caption{Evolution from $z=4$ to $z=0$ of our $(54 ~\rm Mpc/h)^3$ box simulated with $1024^3$ cells and including radiative cooling, AGN feedback and the acceleration efficiency of \citet{kr13}. Top left panel: volume distribution of gas temperature; top right: volume distribution of Mach number; bottom left: $X_{\rm cr}$ as a function of gas overdensity; bottom right: $X_{\rm cr}$ as a function of gas temperature.}
\label{fig:time}
\end{figure*}

In Fig.\ref{fig:time} we show how the distributions of gas temperature, Mach numbers and of $X_{\rm cr}$ evolve with time. We use the $(54 ~\rm Mpc/h)^3$ volume from $z=4$ to $z=0$, employing our full-physics prescription at maximum resolution (run "CUR3\_1024\_c1").\\
Fast motions on large scales induced by structure formation are still limited to a few cluster forming regions, and this produces 
a very steep distribution of Mach number peaking at $M=1.5$ and characterized by  very few  strong, $M \geq 5-10$, shocks. The injection of CRs has just begun at this epoch, and a maximum of $X_{\rm cr} \sim 10$  percent is found only around the first structures, at $\rho \geq 10^3 \rho_{\rm cr}$.\\ 
At later redshifts, the temperature distribution broadens due to the formation of large voids and massive structures. The distribution of Mach numbers flattens considerably with time, and strong shocks become $\sim 10^2-10^3$ times more frequent. These effects combined
lead to the substantial enrichment of CRs across a broad range of gas temperatures and overdensities. 
Interestingly, the function $X_{\rm cr}(\rho)$ is nearly converged from $z=2$ for $\rho>10^2 \rho_{\rm cr}$, while the evolution of $X_{\rm cr}(\rho)$ at lower densities shows a continuous change with redshift. As an example, for $\rho/\rho_{\rm cr} \sim 1$ the pressure support from CRs has increased by nearly 2 orders of magnitude from $z=2$ to $z=0$. This can be explained considering that, after its first assembly, shocks in the virialized ICM are expected to be weak \citep[][]{gb03}. On the other hand, the outer parts of clusters continue to accrete and host shocks that accelerate CRs even at low redshifts. In addition, the drop in the background temperature of the Universe enhances the typical temperature jump experienced by the accreted diffuse gas, thus increasing the average acceleration efficiency at accretion shocks over time, in contrast to what 
is found inside the virial volume of structures. 

\section{Discussion and conclusions}
\label{sec:conclusion}

We simulated the thermal and non-thermal properties of the large-scale structure of the Universe, using new algorithms in the
cosmological code {\enzo}.
Our unigrid simulations provide an extensive survey of models for CRs in cosmological simulations and achieve a high resolution in the outer parts of galaxy clusters and in the most rarefied
cosmic environment.
This study addressed the statistical properties of CR energy in cosmological simulations on 
large scales and at low redshifts ($z \leq 2$). \\
Our parameter study showed:

\begin{itemize}

\item as a rule of thumb, once the spatial resolution is equal or better than $\sim 100  ~ \rm kpc/h$ 
physical effects rather than resolution effects are dominant to set the final level of CRs inside large-scale structures. 

\item At the physical boundaries between the collapsed and the rarefied Universe ($\rho \sim 1-10  ~ \rho_{\rm cr} $ and $T \sim 10^4-10^5  ~\rm K$) the pressure support by CRs accelerated by formation shocks is $\sim 10-20$ percent of the  thermal gas support.  The pressure ratio in this regime appears to have converged with resolution, and is modestly affected (factor $\sim 2$) by the specific implementation of feedback from AGN and SNe and/or by the adoption of different CR acceleration efficiencies. The absolute energy budget of CRs in the case of the \citet{kj07} model is found to be $\sim 50$ percent higher than the budget produced in the \citet{kr13} model in this regime. However,  in this regime the ratio between CR and thermal gas is almost equally affected by uncertainties in the physical prescription for the thermal gas (e.g. in the coupling of radiative cooling and sources of feedback in the ICM) and by the uncertainties on CR physics, if an efficiency of the order of $\sim 10-20$ percent is assumed at $M \geq 5-10$ shocks. 

\item The only exception is given by an early population of SNe (released at $z=2$).  In this case, a significant population of CRs can be advected into the voids, and can have an important dynamical role in the magnetisation of the low-density Universe. On the other hand, the impact of SNe feedback on the budget of CRs within galaxy clusters at low redshift is negligible.

\item At our best resolution ($\approx 52  ~ \rm kpc/h$) the pressure ratio of CRs is  of the order of $\sim 10^{-3}$ inside galaxy clusters, generally within the constraints from $\gamma$-ray observations. Our results in cluster cores still seem to be affected by resolution effects. A further increase in resolution is expected to lower slightly the CRs to gas pressure ratio in cluster centres.  Furthermore, both specific implementations of cooling/AGN feedback and of the injection of CRs can change this value, up to factors of $\sim 5-10$. Future deeper observations of hadronic emission (or upper limits) from the ICM will put stronger constraints on both kinds of mechanisms.

\item The adoption of the \citet{kj07} or \citet{kr13} acceleration efficiencies of CRs at shocks yields no significant difference in average distributions of gas density, temperature and Mach number. However, the average pressure ratio in CRs is lowered by a $\sim 30-60$ percent at all cosmic environments in the \citet{kr13} model, providing less tension with available constraints from the non-detection of hadronic $\gamma$-ray emission from the innermost cluster regions \citep[][]{ack10,2013arXiv1308.6278H,fermi13}.

\item If we neglect the effect of shock re-acceleration of CRs, or if we do not allow acceleration for $M \leq 3-5$ shocks, the pressure ratio of CRs is found to be a factor $\sim 10-30$ lower than the estimates of the
\citet{kj07} and \citet{kr13} models. At low densities, a change in the acceleration efficiency at weak
shocks only affects the normalisation of the dependence of $X_{\rm cr}$ on gas density, while for the
high densities typical of galaxy clusters the slope of this relation is changed also. This suggests a possible
observable test to constrain the functional shape of the $\eta(M)$ acceleration efficiency at weak
shocks.

\item The pressure support from CRs changes significantly from $z=2$ to $z=0$ for $\rho \leq 10^2  ~ \rho_{\rm cr} $. For higher densities the average of $X_{\rm cr}$ remains fairly constant. 

\end{itemize}

\bigskip

In summary, shock-accelerated CRs are expected to be a significant energy component of the Universe, with a maximal dynamical impact in the outer regions of large-scale structures. 
While the exact level of the energy budget in CRs is found to depend on details of the numerical modelling and of the assumed acceleration scenario, the presence of a non-negligible amount of CR energy inside large-scale structures is an unavoidable expectation of DSA during structure formation. 
Yet no detection of hadronic emission from accelerated CRs has been reported so far, limiting the energy budget of CRs to a few percent inside the full virial volume of clusters \citep[][]{dom09,alek10,ack10,arl12,alek12,2013arXiv1308.6278H,fermi13}. 
At present, none of the models for CR acceleration \citep[][]{kj07,kr13} investigated here can be ruled out by our simulations. However, in many cases the implied energy budget of CRs appears close to the available upper limits. Since merger shocks with Mach numbers $M \leq 2-5$ shocks are found to be crucial to set the level of CRs in the innermost regions of clusters, this is the range where the theoretical improvements in our understanding of DSA is expected to increase our understanding of the physics of the ICM.

\bigskip

\section*{acknowledgements}

Computations described in this work were performed using the {\enzo} code (http://enzo-project.org), which is the product of a collaborative effort of scientists at many universities and national laboratories. We gratefully acknowledge the {\enzo} development group for providing extremely helpful and well-maintained on-line documentation and tutorials.\\
We acknowledge PRACE for awarding us access to CURIE-Genci based in France at Bruyeres-le-Chatel. The support of the TGC Hotline from the Centre CEA-DAM Ile de France to the technical work is gratefully acknowledged.
We also acknowledge CSCS-ETH {\footnote{www.cscs.ch}} for the use of the Cray XC30 Piz Daint in order to complete the $2048^3$ run.
F.V. and M.B. acknowledge support from the grant FOR1254 from the Deutsche Forschungsgemeinschaft. 
F.V. and M. B. acknowledge the  usage of computational resources on the JUROPA cluster at the at the Juelich Supercomputing Centre (JSC), under project no. 5018, 5984 and 5056. 
We thank H. Kang, G. Ferini, M. Stubbe for their scientific feedback, and M. Giuffreda and J. Favre for their valuable technical assistence at CSCS.

\bibliographystyle{mnras}
\bibliography{franco}

\end{document}